\documentclass[useAMS,usenatbib]{mn2e} 
\usepackage{aas_macros}
\usepackage{graphics}
\usepackage{times}
\usepackage{amsmath,amssymb}
\usepackage{bm}
\usepackage{color}
\usepackage{hyperref}
\usepackage{fixltx2e}

\def\be{\begin{equation}}
\def\ee{\end{equation}}
\def\ba{\begin{eqnarray}}
\def\ea{\end{eqnarray}}

\def\lsim{\mathrel{\rlap{\lower4pt\hbox{\hskip1pt$\sim$}}
    \raise1pt\hbox{$<$}}}                
\def\gsim{\mathrel{\rlap{\lower4pt\hbox{\hskip1pt$\sim$}}
    \raise1pt\hbox{$>$}}}

\pagerange{\pageref{firstpage}--\pageref{lastpage}}
\pubyear{2015}

\begin{document}

\label{firstpage}

\title[Measuring cosmic shear and birefringence]{Measuring cosmic shear and birefringence using resolved radio sources}
\author[Lee Whittaker, Richard A. Battye \& Michael L. Brown]{Lee Whittaker, Richard A. Battye \& Michael L. Brown\\ Jodrell Bank Centre for Astrophysics,
  School of Physics and Astronomy, University of Manchester, Oxford
  Road, Manchester M13 9PL} \date{\today}

\maketitle

\begin{abstract}
We develop a new method of extracting simultaneous measurements of weak lensing shear and a local rotation of the plane of polarization using observations of resolved radio sources. {We show that the direction of polarization is statistically linked with that of the gradient of the total intensity field, and this provides the basis of our method}. Using a number of sources spread over the sky, this method allows constraints to be placed on cosmic shear and birefringence, and it can be applied to any resolved radio sources for which such a correlation exists. Assuming that the rotation and shear are constant across the source, we use this relationship to construct a quadratic estimator and investigate its properties using simulated observations. We develop a calibration scheme using simulations based on the observed images to mitigate a bias which occurs in the presence of measurement errors and an astrophysical scatter on the polarization. The method is applied directly to archival data of radio galaxies where we measure a mean rotation signal of $\left<\omega\right>=-2.03^{\circ}\pm0.75^{\circ}$ and an average shear compatible with zero using 30 reliable sources. This level of constraint on an overall rotation is comparable with current leading constraints from CMB experiments and is expected to increase by at least an order of magnitude with future high precision radio surveys, such as those performed by the SKA. We also measure the shear and rotation two-point correlation functions and estimate the number of sources required to detect shear and rotation correlations in future surveys.
\end{abstract}

\begin{keywords}
gravitational lensing: weak - methods: analytical - methods: statistical - cosmology: theory
\end{keywords}
\section{Introduction}
\label{sec:intro}
Cosmic shear is the coherent distortion of the shapes of background galaxies by the large scale distribution of foreground matter. The measurement of the cosmic shear signal is now considered a powerful probe of cosmology due to its sensitivity to the integrated mass along the line of sight. In recent years, weak lensing surveys, such as the Canada-France-Hawaii Telescope Lensing Survey (CFHTLenS) \citep{heymans13}, the Dark Energy Survey Science Verification (DES SV) \citep{abbott15}, and using $\sim$450 sq degrees of the Kilo-Degree Survey (KiDS-450) \citep{hildebrandt16}, have placed constraints on the total matter density parameter and the amplitude of the linear matter power spectrum, and future surveys, such as those performed by \emph{Euclid}\footnote{\href{http://sci.esa.int/euclid}{http://sci.esa.int/euclid}}, the Large Synoptic Survey Telescope\footnote{\href{https://www.lsst.org}{https://www.lsst.org}} (LSST), and the Wide-Field Infrared Survey Telescope\footnote{\href{http://wfirst.gsfc.nasa.gov}{http://wfirst.gsfc.nasa.gov}} (WFIRST) aim to place tight constraints on the dark energy equation of state (e.g. \citealt{schrabback10, kilbinger13}).

The intrinsic shapes of galaxies are not known, therefore the standard method for measuring cosmic shear assumes that the galaxies are intrinsically randomly orientated. Under this assumption, the measured shape of a galaxy provides an unbiased estimate of the shear at that position on the sky. The intrinsic randomness in galaxy shapes contributes a shot noise error to estimates of the shear. Furthermore, the assumption that the galaxies are randomly orientated is expected to break down for galaxies that share an evolutionary history. For such galaxies, tidal effects can cause galaxies to become intrinsically aligned with the local large-scale structure (\citealt{heavens00, croft00, crittenden01, catelan01}). This alignment biases shear estimates and consequently estimates of the cosmological parameters \citep{heymans13}. For a given scale, errors on estimates of the cosmic shear signal, resulting from shape noise and errors on the shape measurements, are typically beaten down by averaging over the shapes of a large number of galaxies. The number density of source galaxies in a survey therefore plays a key role in determining the power with which the survey can constrain cosmology. Current state-of-the-art surveys are in the optical waveband and are detecting galaxies at a number density of $\sim$$10\,\mathrm{arcmin}^{-2}$ over an area of $\sim$$5,000\,\mathrm{deg}^2$. Euclid aims to push the number density of observed galaxies up to $30\,\mathrm{arcmin}^{-2}$ over an area of $\sim$$15,000\,\mathrm{deg}^2$.

Detections of cosmic shear have also been made the in radio band \citep{chang04}, and by cross-correlating optical and radio observations using the SDSS and the VLA FIRST survey \citep{demetroullas16}. However, radio weak lensing is currently uncompetitive with optical weak lensing due to the low source density. This position is likely to change in the future with the advent of a whole new raft of facilities. The SKA, for example, is expected to perform surveys competitive with optical surveys, achieving number densities $\sim$$10\,\mathrm{arcmin}^{-2}$ over an area of $\sim$$30,000\,\mathrm{deg}^2$ \citep{brown15}.

Recent work by \cite{brown11a}  and \cite{whittaker15a} proposed statistical approaches for using polarization information from radio surveys to reduce the effects of shot noise and intrinsic alignments. These ideas rely on the hypothesis that the integrated polarization position angle is anti-correlated with the structural position angle for star-forming galaxies, which will be the dominant population at the flux densities needed to achieve source densities $\sim$$10\,\mathrm{arcmin}^{-2}$. Such a relationship has been established in the local Universe {\citep{stil09}}, and this is expected to be universal at some level, providing unbiased estimates of the intrinsic position angles of the galaxies.

In this paper, we develop a related but different approach to detecting weak lensing using radio observations. We propose to use high signal-to-noise and high resolution observations to measure the local shear field in the direction of individual resolved radio galaxies. We show that there exists a statistical alignment between the gradient of the total intensity field and the polarization position angle for a sample of well studied galaxies. This alignment will be broken in the presence of a shear and/or rotation of the plane of polarization, and by mapping the gradient and polarization position angles across a well resolved source, one can use this break in alignment to estimate the signal. Using a large number of source galaxies ($\sim$$10^5$), the shear field can then be mapped over a significant portion of the sky. The ultimate aim is to apply this technique to data from the SKA, but proof of principle on surveys of smaller areas should be possible with the SKA progenitors. In fact, a similar method has already been used by \cite{kronberg91}, \cite{kronberg96}, and \cite{burns04} to establish the existence of a lensing galaxy in front of the giant radio source 3C9. Their approach was to search for the effects of individual objects. What we describe here is more general and aims to detect the extended distribution of matter across a large area of the sky.

In addition to measuring cosmic shear, an estimate of the rotation signal can be used to probe cosmic birefringence. Cosmic birefringence is the hypothesized rotation of the plane of polarization of electromagnetic radiation due to a parity violating modification of Maxwell's laws (e.g. { (\citealt{carroll90,harari92,colladay98,kosteleck09})}). A rotation of the plane of polarization does not change the observed shape of a source which makes this effect different to that of a shear signal. This difference between the effects of shear and rotation allows the two signals to be separated for a sufficiently resolved source. There has been a previous claim of a detection of cosmic birefringence using the integrated polarization from extra-galactic radio sources \citep{nodland97}; however, this claim has since been shown to be statistically insignificant (\citealt{carroll97,loredo97}), and today the strongest constraints on an overall global rotation come from cosmic microwave background (CMB) data {(\citealt{alighieri14,ade15b,gruppuso16})}. Despite the strong constraints that come from the CMB, there is interest in a spatially correlated rotation {(e.g. \citealt{caldwell11,lee15,namikawa17})}, and a large survey of sources could allow this to be measured.

In Section \ref{sec:estimator} we derive the estimator and discuss the assumption that the linear polarization vector is aligned with the gradient of the total intensity distribution. In Section \ref{sec:simple_sims} we simulate observations of Fanaroff and Riley Class II (FRII)-like sources using sums of Gaussian profiles and use these simulations to test the performance of our estimator; although we note that the estimator should work for any type of source which is resolved. We also develop a method to correct for the biases that are introduced in the presence of measurement errors and an astrophysical scatter between the polarization vector and the gradient.

In Section \ref{sec:character_data} we describe the VLA FRII data used to measure the shear and rotation two point correlation functions in Section \ref{sec:results}. Finally, in Section \ref{sec:future} we constrain the number of sources required for a survey performed using the SKA to make a detection of the shear signal using this approach. We conclude in Section \ref{sec:conclude}. 

\section{Construction of the estimator}
\label{sec:estimator}
The effect of gravitational lensing on any Stokes parameter distribution can be described by $S_{\mathrm{obs}}\left(\bm{x}\right)=S\left(\bm{x}'\right)=S\left(\bm{x}-\bm{\nabla}\psi\right)$, where $S=I,Q,U$, and $\psi$ is the lensing potential. $I\left(\bm{x}\right)$ describes the total intensity distribution of the source, and $Q\left(\bm{x}\right)$ and $U\left(\bm{x}\right)$ describe the linear polarization. $S_{\mathrm{obs}}$ is the observed distribution of the Stokes parameters in the image plane and $S$ is the intrinsic distribution in the source plane. Assuming that the properties of the lens vary slowly compared to the size of the source, we can perform a Taylor expansion on $\bm{\nabla}\psi$, so that
\begin{equation}
x_i'=x_i-\nabla\psi=x_i-\nabla_i\psi|_{0}-x_j\nabla_j\nabla_i\psi|_{0},
\end{equation}
which can be written as
\begin{equation}
\bm{x}'=\bm{d}+\mathbf{A}\bm{x}.
\end{equation}
The term $\bm{d}=-\bm{\nabla}\psi|_0$ is an undetectable offset that, for simplicity, we assume to be zero for the remainder of this paper. The matrix $A_{i,j}=\delta_{i,j}-\nabla_i\nabla_j\psi$ is the Jacobian of the transformation. The convergence is defined in terms of the lensing potential as $\kappa=\frac{1}{2}\nabla^2\psi$, and the shear, $\bm{\gamma}$, is defined as
\begin{align}\label{eq:def_shear}
\gamma_1=&\frac{1}{2}\left(\nabla_1^2-\nabla_2^2\right)\psi,\nonumber\\
\gamma_2=&\nabla_1\nabla_2\psi,
\end{align}
so that the Jacobian matrix becomes
\begin{equation}\label{eq:lens_A}
\mathbf{A}=\left(
\begin{array}{cc}
1-\kappa-\gamma_1 & -\gamma_2 \\
-\gamma_2 & 1-\kappa+\gamma_1
\end{array} \right).
\end{equation}

In addition to the effects of the shear, an overall rotation of the Stokes distributions can be described by defining the rotation matrix
\begin{equation}\label{eq:rot_R}
\mathbf{R}=\left(
\begin{array}{cc}
\cos\omega & -\sin\omega \\
\sin\omega & \cos\omega
\end{array} \right),
\end{equation}
where $\omega$ is the rotation angle of the plane of polarization. Assuming a zero shear signal and a non-zero rotation, a Stokes parameter distribution is transformed as $S_{\mathrm{obs}}\left(\bm{x}\right)=S\left(\bm{x}'\right)=S\left(\mathbf{R}^{\mathrm{T}}\bm{x}\right)$. If we now assume a non-zero shear signal and rotation before lensing, the corresponding coordinate transformation is
\begin{equation}\label{eq:rot_before}
\bm{x}'=\mathbf{R}^{\mathrm{T}}\mathbf{A}\bm{x},
\end{equation}
whereas if the rotation occurs after lensing,
\begin{equation}\label{eq:rot_after}
\bm{x}'=\mathbf{A}\mathbf{R}^{\mathrm{T}}\bm{x}.
\end{equation}
To first order in $\omega$, $\kappa$ and $\bm{\gamma}$, it can be shown that both of the transformations given in equations (\ref{eq:rot_before}) and (\ref{eq:rot_after}) are identical. That is, for $\omega,\kappa,\bm{\gamma}\ll1$,
\begin{equation}\label{eq:first_order_rot}
\tilde{\mathbf{A}}\bm{x}\approx\mathbf{R}^{\mathrm{T}}\mathbf{A}\bm{x}\approx\mathbf{A}\mathbf{R}^{\mathrm{T}}\bm{x},
\end{equation}
with
\begin{equation}\label{eq:A_w}
\tilde{\mathbf{A}}=\left(
\begin{array}{cc}
1-\kappa-\gamma_1 & -\gamma_2+\omega \\
-\gamma_2-\omega & 1-\kappa+\gamma_1
\end{array} \right).
\end{equation}
Assuming both a rotation and lensing, we can therefore approximate the transformation of a distribution of Stokes parameters to first order in $\bm{\gamma}$ and $\omega$ as $S_{\mathrm{obs}}\left(\bm{x}\right)=S\left(\tilde{\mathbf{A}}\bm{x}\right)$. For convenience, we now redefine the matrix $\mathbf{A}$ as $\mathbf{A}\equiv\tilde{\mathbf{A}}$.
 
We can express Stokes $Q$ and $U$ in polar form:
\begin{align}\label{eq:Q_U_polar}
Q\left(\bm{x}\right)=&P\left(\bm{x}\right)\cos\left[2\alpha\left(\bm{x}\right)\right],\nonumber\\
U\left(\bm{x}\right)=&P\left(\bm{x}\right)\sin\left[2\alpha\left(\bm{x}\right)\right],
\end{align}
where $P\left(\bm{x}\right)$ is the polarized intensity distribution and $\alpha\left(\bm{x}\right)$ is the distribution of polarization position angles. The factor of two is included as the linear polarization is described by a spin-2 vector, which is invariant under rotations of $180^{\circ}$. The fact that the distributions of Stokes $Q$ and $U$ transform as $Q_{\mathrm{obs}}\left(\bm{x}\right)=Q\left(\bm{x}'\right)$ and $U_{\mathrm{obs}}\left(\bm{x}\right)=U\left(\bm{x}'\right)$ implies that
\begin{equation}\label{eq:pol_a_relate}
\hat{n}_{P_{\mathrm{obs}}}\left(\bm{x}\right)\equiv\left(
\begin{array}{c}
\cos\left(\alpha_{\mathrm{obs}}\left(\bm{x}\right)\right) \\
\sin\left(\alpha_{\mathrm{obs}}\left(\bm{x}\right)\right)
\end{array} \right)=\hat{n}_{P}\left(\bm{x}'\right)\equiv\left(
\begin{array}{c}
\cos\left(\alpha\left(\bm{x}'\right)\right) \\
\sin\left(\alpha\left(\bm{x}'\right)\right)
\end{array} \right).
\end{equation}.

It can be shown that the effect of shear and rotation on the gradient of $I_{\mathrm{obs}}\left(\bm{x}\right)$ is
\begin{equation}\label{eq:lens_gradI}
\bm{\nabla}_xI_{\mathrm{obs}}\left(\bm{x}\right)=\mathbf{A}^{\mathrm{T}}\bm{\nabla}_{x'}I\left(\bm{x}'\right).
\end{equation}
Defining $\beta\left(\bm{x}\right)$ as the local direction of $\bm{\nabla}_xI_{\mathrm{obs}}\left(\bm{x}\right)$, we have
\begin{equation}
\hat{n}_I\left(\bm{x}\right)\equiv\frac{\bm{\nabla}_xI\left(\bm{x}\right)}{\left|\bm{\nabla}_xI\left(\bm{x}\right)\right|}=
\left(
\begin{array}{c}
\cos\left(\beta\left(\bm{x}\right)\right) \\
\sin\left(\beta\left(\bm{x}\right)\right)
\end{array} \right).
\end{equation}

 Let us assume, for the moment, that the linear polarization is perfectly aligned with the gradient of the intensity distribution
\begin{equation}\label{eq:assumption}
\hat{n}_I\left(\bm{x}\right)=\pm\hat{n}_{P}\left(\bm{x}\right),
\end{equation}
where the plus or minus accounts for the quadrant of the gradient vector as $\alpha$ is only defined on the range $-90^{\circ}\le\alpha<90^{\circ}$, whereas $\beta$ is defined over $-180^{\circ}\le\beta<180^{\circ}$.

For this ideal case, we can relate the gradient of the observed intensity distribution to the measured polarization position angles,
\begin{equation}\label{eq:pre-exp}
\hat{n}_{I_{\mathrm{obs}}}\left(\bm{x}\right)=\frac{\bm{\nabla}_xI_{\mathrm{obs}}\left(\bm{x}\right)}{\left|\bm{\nabla}_xI_{\mathrm{obs}}\left(\bm{x}\right)\right|}=
K\mathbf{A}^{\mathrm{T}}\hat{n}_{P_{\mathrm{obs}}}\left(\bm{x}\right),
\end{equation}
with
\begin{equation}\label{K}
K=\pm\frac{\left|\bm{\nabla}_{x'}I\left(\bm{x}'\right)\right|}{\left|\bm{\nabla}_xI_{\mathrm{obs}}\left(\bm{x}\right)\right|},
\end{equation}
which contains the unobservable $\left|\bm{\nabla}_{x'}I\left(\bm{x}'\right)\right|$. 

Expanding equation (\ref{eq:pre-exp}) we have
\begin{align}\label{eq:sim_eqs_est}
\cos\beta_{\mathrm{obs}}\left(\bm{x}\right)=&K\left(A_{11}\cos\alpha_{\mathrm{obs}}\left(\bm{x}\right)+A_{21}\sin\alpha_{\mathrm{obs}}\left(\bm{x}\right)\right),\nonumber\\
\sin\beta_{\mathrm{obs}}\left(\bm{x}\right)=&K\left(A_{22}\sin\alpha_{\mathrm{obs}}\left(\bm{x}\right)+A_{12}\cos\alpha_{\mathrm{obs}}\left(\bm{x}\right)\right),
\end{align}
and then taking the ratio of the two equations in equation (\ref{eq:sim_eqs_est}), we can eliminate $K$, giving
\begin{align}\label{eq:elim_K}
\cos\beta_{\mathrm{obs}}&\left(\bm{x}\right)\left(A_{22}\sin\alpha_{\mathrm{obs}}\left(\bm{x}\right)+A_{12}\cos\alpha_{\mathrm{obs}}\left(\bm{x}\right)\right)\nonumber\\
&=\sin\beta_{\mathrm{obs}}\left(\bm{x}\right)\left(A_{11}\cos\alpha_{\mathrm{obs}}\left(\bm{x}\right)+A_{21}\sin\alpha_{\mathrm{obs}}\left(\bm{x}\right)\right).
\end{align}
If we now consider a source image consisting of $N$ pixels with a signal-to-noise above a desired threshold level and with a measurement of $\alpha_{\mathrm{obs}}$ and $\beta_{\mathrm{obs}}$ for each pixel, we can define the $\chi^2$
\begin{align}\label{eq:chi2}
\chi^2=&\sum_{k=1}^N\biggl[\cos\beta_{\mathrm{obs}}^{(k)}\left(A_{22}\sin\alpha_{\mathrm{obs}}^{(k)}+A_{12}\cos\alpha_{\mathrm{obs}}^{(k)}\right)\nonumber\\
&-\sin\beta_{\mathrm{obs}}^{(k)}\left(A_{11}\cos\alpha_{\mathrm{obs}}^{(k)}+A_{21}\sin\alpha_{\mathrm{obs}}^{(k)}\right)\biggr]^2.
\end{align}

Minimizing this $\chi^2$ with respect to $\bm{\gamma}$ and $\omega$, we recover the estimator
\begin{equation}\label{eq:estimator}
\hat{\bm{\Gamma}}=\mathbf{B}^{-1}\bm{d},
\end{equation}
where
\begin{equation}\label{eq:Gamma}
\bm{\Gamma}=\left(
\begin{array}{c}
\omega \\
\gamma_1 \\
\gamma_2
\end{array} \right).
\end{equation}
This estimator is similar in form to the shear estimator derived by \cite{brown11a} which also includes polarization information.

The matrix $\mathbf{B}$ is given as
\begin{equation}\label{eq:mat_B}
\mathbf{B}=\frac{1}{N}\sum_{k=1}^N\left(
\begin{array}{ccc}
1+C_-^{(k)} & S_{\alpha}^{(k)}+S_{\beta}^{(k)} & -\left(C_{\alpha}^{(k)}+C_{\beta}^{(k)}\right) \\
S_{\alpha}^{(k)}+S_{\beta}^{(k)} & 1-C_+^{(k)} & -S_+^{(k)} \\
-\left(C_{\alpha}^{(k)}+C_{\beta}^{(k)}\right) & -S_+^{(k)} & 1+C_+^{(k)}
\end{array} \right),
\end{equation}
and the vector $\bm{d}$ is
\begin{equation}\label{eq:vec_d}
\bm{d}=\frac{1}{N}\sum_{k=1}^N\left(
\begin{array}{c}
-S_-^{(k)} \\
C_{\alpha}^{(k)}-C_{\beta}^{(k)} \\
S_{\alpha}^{(k)}-S_{\beta}^{(k)}
\end{array} \right),
\end{equation}
where
\begin{align}
C_{\alpha}^{(k)}=&\cos\left(2\alpha_{\mathrm{obs}}^{(k)}\right),\quad C_{\beta}^{(k)}=\cos\left(2\beta_{\mathrm{obs}}^{(k)}\right),\nonumber\\
S_{\alpha}^{(k)}=&\sin\left(2\alpha_{\mathrm{obs}}^{(k)}\right),\quad S_{\beta}^{(k)}=\sin\left(2\beta_{\mathrm{obs}}^{(k)}\right),\nonumber\\
C_-^{(k)}=&\cos\left[2\left(\alpha_{\mathrm{obs}}^{(k)}-\beta_{\mathrm{obs}}^{(k)}\right)\right],\quad C_+^{(k)}=\cos\left[2\left(\alpha_{\mathrm{obs}}^{(k)}+\beta_{\mathrm{obs}}^{(k)}\right)\right],\nonumber\\
S_-^{(k)}=&\sin\left[2\left(\alpha_{\mathrm{obs}}^{(k)}-\beta_{\mathrm{obs}}^{(k)}\right)\right],\quad S_+^{(k)}=\sin\left[2\left(\alpha_{\mathrm{obs}}^{(k)}+\beta_{\mathrm{obs}}^{(k)}\right)\right].
\end{align}

So far we have assumed that the correlation between $\alpha$ and $\beta$ is exact, that is, $\alpha-\beta$ is zero at all points in space. However, while there are physical reasons to expect a close correlation between the two, it is unlikely to be perfect. We will model the deviations from a perfect correlation by Gaussian random fluctuations added to each of the Stokes parameters, $Q$ and $U$, with standard deviations, $\sigma_Q$ and $\sigma_U$, respectively, which will of course be an approximation to what is a highly complicated situation. 

In order to understand the expected distribution of $\alpha-\beta$, let us first consider the angle distribution created by random fluctuations about zero with standard deviations in Stokes parameters $Q=P\cos\theta$ and $U=P\sin\theta$ where $-\pi\le\theta<\pi$ ignoring for the moment the spin-two nature of the polarization. One can show that this is given by 
\begin{equation}
{\cal P}(\theta)d\theta={\sinh\left(2\eta^2\right)\,d\theta\over 2\pi\left[\cosh\left(2\eta^2\right)-\cos 2\theta\right]}\,,
\label{projnorm}
\end{equation}
where
\begin{equation}
\cosh\left(2\eta^2\right)={\sigma_Q^2+\sigma_U^2\over |\sigma_Q^2-\sigma_U^2|}\,,\quad
\sinh\left(2\eta^2\right)={2\sigma_Q\sigma_U\over |\sigma_Q^2-\sigma_U^2|}\,.
\end{equation}
This is a 2D projected Normal distribution $PN_2({\bf 0},\Sigma)$ with zero mean and covariance matrix $\Sigma={\rm diag}(\sigma_Q^2,\sigma_U^2)$. There is a well known equivalence between such a distribution and the Wrapped Cauchy distribution $WC(0,\rho)$. In particular, if $\theta\sim PN_2({\bf 0},\Sigma)$ then $2\theta\sim WC(0,\rho)$ where 
\begin{equation}
\cosh\left(2\eta^2\right)={1+\rho^2\over 2\rho}\,,\quad\sinh\left(2\eta^2\right)={1-\rho^2\over 2\rho}\,,
\end{equation}
and the parameter $\eta$ describes the dispersion of $\theta$ in units of radians.

However, the situation we are interested in is slightly different to this. In particular we are dealing with a position angle spin-two polarization position angle, $-\pi/2\le\alpha<\pi/2$, and a real space position angle, $-\pi\le\beta<\pi$, but any deviations from a perfect correlation will be modelled by standard deviations in the Stokes parameters as above. Taking this into account one can model $\alpha-\beta$ using the distribution in equation (\ref{projnorm}) above with the transformation $\theta=\alpha-\beta$ for $-\pi/2\le\theta<\pi/2$, $\alpha-\beta=\theta+\pi$ for $-\pi\le\theta<-\pi/2$ and $\alpha-\beta=\theta-\pi$ for $\pi/2\le\theta<\pi$. Hence, we deduce that 
\begin{equation}\label{eq:wc}
{\cal P}(\alpha-\beta)d(\alpha-\beta)={\sinh\left(2\eta^2\right)\,d(\alpha-\beta)\over \pi(\cosh\left(2\eta^2\right)-\cos[2(\alpha-\beta)]}\,,
\end{equation}
for $-\pi/2\le\alpha-\beta<\pi/2$. We find that this distribution fits the data well and we use this distribution in the subsequent analysis to model the deviations form a perfect correlation in the simulations used to estimate errors in our birefringence/shear estimates. {It should be noted that there is no justification from first principles for modeling the distributions of $Q$ and $U$ as Gaussian; however, we find that this assumption provides a good fit to the distribution of measured $\alpha-\beta$ via equation (\ref{eq:wc}). There will be a random scatter in the orientation of the polarization vectors due to turbulence in the magnetic fields, and hence if one assumes the central limit theorem, one might expect the observed $Q$ and $U$ to be approximately Gaussian distributed.}

\section{Simple test simulations}
\label{sec:simple_sims}
We have tested the estimator given in equation (\ref{eq:Gamma}) using simulations where we assumed a simple model for a jet/lobe dominated radio source. This model consists of three Gaussian distributions - a highly elliptical Gaussian for the jet, and one circular Gaussian for each of the two lobes. Using this toy model, we created noisy $I$, $Q$ and $U$ maps which included both a shear and rotation signal. We do not claim that this is an accurate representation of radio sources, just that it is similar to sources we will study in subsequent sections. We then tested the performance of the estimator by recovering estimates of the input signals from the simulated maps.

For this simple model, the total intensity distribution for the source is
\begin{align}\label{eq:sim_gauss_image}
I_{\mathrm{obs}}\left(\bm{x}\right)=&I_0\left(\bm{x}\right)+I_1\left(\bm{x}\right)+I_2\left(\bm{x}\right),\nonumber\\
I_0\left(\bm{x}\right)=&\bar{I}_0\exp\left[-\frac{1}{2}\left(\mathbf{A}\bm{x}-\bar{\bm{x}}_0\right)^{\mathrm{T}}\mathbf{M}_0\left(\mathbf{A}\bm{x}-\bar{\bm{x}}_0\right)\right],\nonumber\\
I_1\left(\bm{x}\right)=&\bar{I}_1\exp\left[-\frac{1}{2}\left(\mathbf{A}\bm{x}-\bar{\bm{x}}_1\right)^{\mathrm{T}}\mathbf{M}_1\left(\mathbf{A}\bm{x}-\bar{\bm{x}}_1\right)\right],\nonumber\\
I_2\left(\bm{x}\right)=&\bar{I}_2\exp\left[-\frac{1}{2}\left(\mathbf{A}\bm{x}-\bar{\bm{x}}_2\right)^{\mathrm{T}}\mathbf{M}_2\left(\mathbf{A}\bm{x}-\bar{\bm{x}}_2\right)\right],
\end{align}
where the subscripts denote the individual components of the simulated image, $\bar{I}_n$ is the peak intensity of component $n$, and $\bar{\bm{x}}_n$ is the centroid of component $n$. For our simulations, the zeroth component was used to model the jet and the first and second components were used to model the lobes of the source. The matrices $\mathbf{M}_n$ are quadrupole matrices describing the shapes of the components, with
\begin{figure}
\includegraphics{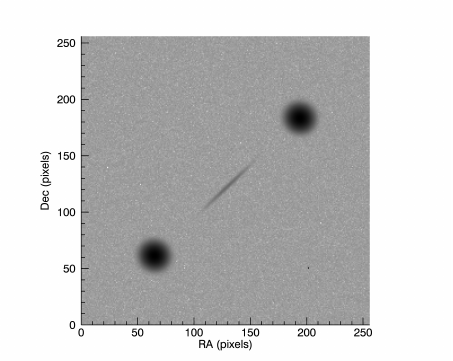}
\caption{An example of the simulated high signal-to-noise total intensity images used to test the estimator. The simulation is constructed from three Gaussian components using equation (\ref{eq:sim_gauss_image}). Equivalent maps were also constructed for the Stokes $Q$ and $U$ distributions.}
\label{fig:sim_gauss}
\end{figure}
\begin{align}\label{eq:components_of_M}
M_{11}^{(n)}=&\frac{\cos^2\theta_n}{a_n^2}+\frac{\sin^2\theta_n}{b_n^2},\nonumber\\
M_{22}^{(n)}=&\frac{\sin^2\theta_n}{a_n^2}+\frac{\cos^2\theta_n}{b_n^2},\nonumber\\
M_{12}^{(n)}=&M_{21}^{(n)}=\left(\frac{1}{a_n^2}-\frac{1}{b_n^2}\right)\sin\theta_n\cos\theta_n.
\end{align}
Here, $a_n$ and $b_n$ are the major and minor axes of the components respectively, and $\theta_n$ are the corresponding position angles. An example of the simulated sources used to test the method is shown in Figure \ref{fig:sim_gauss}. This image consists of a $256\,\mathrm{pixel}\times256\,\mathrm{pixel}$ grid. We set $a_0=15.0$, $b_0=1.0$, $\theta_0=45.0^{\circ}$, $a_1=b_1=a_2=b_2=5.0$ and $\theta_1=\theta_2=0.0$. The centroids of the components are $\bar{\bm{x}}_0=[128.0, 128.0]$, $\bar{\bm{x}}_1=[64.0,64.0]$, and $\bar{\bm{x}}_2=[192.0, 192.0]$. The background noise was assumed to be Gaussian, and we set the dispersion to unity.
\subsection{High signal-to-noise tests}
\label{sub:high_snr}

We began by testing the method on high signal-to-noise realizations of the source. For the initial tests, we set the peak intensities to $\bar{I}_0=20.0$, $\bar{I}_1=\bar{I}_2=1000.0$ to give a significant number of high signal-to-noise pixels.

To complete our simulations, we simulated images of the Stokes $Q$ and $U$ distributions by assuming that the polarized intensity distribution was identical to the total intensity distribution; that is, we initially set $P\left(\bm{x}\right)=I\left(\bm{x}\right)$. The polarization position angles were derived analytically using the assumption of equation (\ref{eq:assumption}). Letting $\bm{z}_n=\mathbf{A}\bm{x}-\bar{\bm{x}}_n$, the unit vector for the polarization is
\begin{align}\label{eq:unit_vector}
&\bm{n}_P\left(\bm{x}\right)=\frac{\bm{\nabla}_xI\left(\bm{x}\right)}{\left|\bm{\nabla}_xI\left(\bm{x}\right)\right|}\nonumber\\
&=-\frac{\mathbf{M}_0\left(\bm{z}_0\right)I_0\left(\bm{x}\right)+\mathbf{M}_1\left(\bm{z}_1\right)I_1\left(\bm{x}\right)+\mathbf{M}_2\left(\bm{z}_2\right)I_2\left(\bm{x}\right)}{\left|\mathbf{M}_0\left(\bm{z}_0\right)I_0\left(\bm{x}\right)+\mathbf{M}_1\left(\bm{z}_1\right)I_1\left(\bm{x}\right)+\mathbf{M}_2\left(\bm{z}_2\right)I_2\left(\bm{x}\right)\right|}.
\end{align}
We created images of $Q\left(\bm{x}\right)$ and $U\left(\bm{x}\right)$ at the same pixel scale as $I\left(\bm{x}\right)$.

Once these intrinsic $I$, $Q$ and $U$ maps had been constructed, a constant shear and rotation signal were applied. For these tests, we assumed an input shear of $\gamma_1=0.01$ and $\gamma_2=-0.02$, and included a rotation of $\omega=-1.0^{\circ}$. 

 For these first simple simulations, we ignored the effects of beam convolution and assumed Gaussian errors on the components of the gradient vector, with the dispersion equal to $\left|\bm{\nabla}I_{\mathrm{obs}}\right|/\sqrt{2}$; this gives an error on the position angle that is dependent on the modulus of the gradient and is consistent with using the finite differencing method discussed later in this section but avoids the systematic effects of pixelization. The polarization is assumed to be perfectly aligned with the gradient, and the errors on $Q_{\mathrm{obs}}$ and $U_{\mathrm{obs}}$ were assumed to be Gaussian with zero mean and the dispersion equal to unity.

For a constant level of background noise, the error on the measurement of $\alpha_{\mathrm{obs}}$ depends on the polarized intensity, $P_{\mathrm{obs}}$. In order to avoid complications when calibrating for noise bias -- discussed later in this section -- we chose to make a very high signal-to-noise cut on the Stokes $Q$ and $U$ maps. For these tests, we concentrated our analysis on regions of the images where $P_{\mathrm{obs}}\left(\bm{x}\right)>30\sigma_{p}$, where $\sigma_{p}$ is the dispersion of background noise in the $Q$ and $U$ maps (for these tests $\sigma_{p}=1$, as discussed above). Similarly, the error on the measurement of $\beta_{\mathrm{obs}}$ depends on the magnitude of the gradient, $\left|\bm{\nabla}I_{\mathrm{obs}}\right|$. We therefore made a signal-to-noise cut so that we consider only the regions where $\left|\bm{\nabla}I_{\mathrm{obs}}\right|\left(\bm{x}\right)>30\sigma_{\mathrm{b}}/\sqrt{2}$ , where $\sigma_{\mathrm{b}}$ is the background noise level for the total intensity image. The factor of $\sqrt{2}$ reflects the fact that each component of the gradient calculated using the finite differencing method discussed in the following subsection considers the difference between two pixel values separated by a distance of two pixels, each with an independent noise realization. The number of pixels above the signal-to-noise threshold in the noise free maps was $N_{\mathrm{pix}}=998$; however, in all tests conducted in this paper, the signal-to-noise cuts were made on the noisy images, and hence the precise number of pixels considered in each realization varies somewhat due to noise.

\begin{figure*}
\begin{minipage}{6in}
\includegraphics{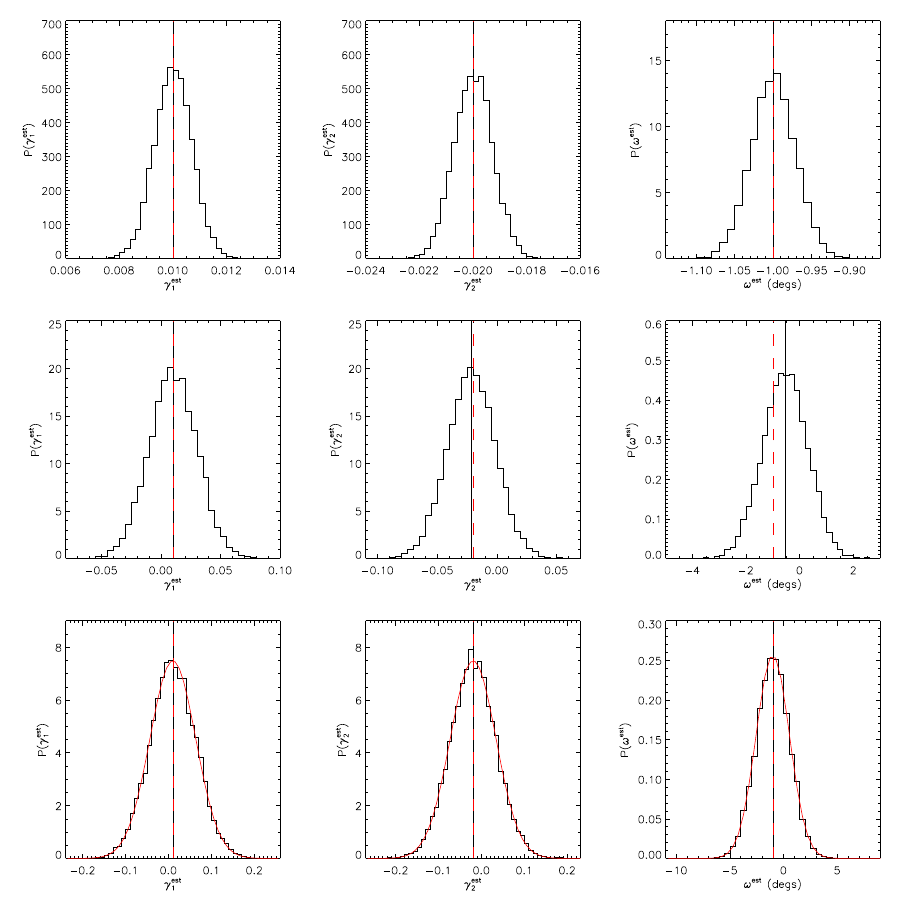}
\caption{Recovered estimates of the shear and rotation from $10^4$ noise realizations. The top panels are the results for the high-signal-to-noise tests described in Section \ref{sub:high_snr}, which appears to give unbiased reconstructions of $\bm{\gamma}$ and $\omega$. The middle row of panels are the results for the realistic noise tests, without noise bias correction, discussed in Section \ref{sub:real_snr}, for which there is a bias in the reconstruction, most obvious for $\omega$. The bottom row shows the same set of realizations as in the middle panels but with the bias correction applied. In each panel the vertical black line shows the mean recovered estimate and the dashed vertical red line shows the input signal. The red curves in the bottom panels are Gaussian distributions with the means equal to the mean recovered estimates, and the dispersions calculated using equation (\ref{eq:errors_circ}).}
\label{fig:all_simple_sims}
\end{minipage}
\end{figure*}

We recovered shear estimates from $10^4$ noise realizations, with the results presented in the top three panels of Figure \ref{fig:all_simple_sims}. The mean estimates were $\left<\hat{\gamma}_1\right>=0.0100$, $\left<\hat{\gamma}_2\right>=-0.0200$, $\left<\hat{\omega}\right>=-1.00^{\circ}$. The error bars on the mean estimates were $<0.1\%$ of the recovered mean signal in all three cases, and the mean estimates were all consistent with the input signals. This shows that the estimator in Section \ref{sec:estimator} is unbiased for this simple test.

\subsection{Tests including realistic noise}
\label{sub:real_snr}
Next we tested the method on more realistic simulations, which included realistic signal-to-noise levels and considered an approach to measuring the total intensity gradient which one might use in reality.

The signal-to-noise cut on the polarized intensity for these tests was made at $P_{\mathrm{obs}}=5\sigma_{P}$, and the signal-to-noise cut on the gradient was made at $\left|\bm{\nabla}I_{\mathrm{obs}}\right|=5\sigma_{\mathrm{b}}/\sqrt{2}$. This is identical to the signal-to-noise cut made on the VLA data discussed in Section \ref{sec:character_data}. As explained in the previous section, errors on the measurements of $\beta$ depend on $\left|\bm{\nabla}I_{\mathrm{obs}}\right|$. However, the signal-to-noise cut on $\left|\bm{\nabla}I_{\mathrm{obs}}\right|$ selects regions of the total intensity images where the contours are closely packed and can ignore regions where the signal-to-noise of the intensity images are high, such as the centres of the sources. The simulated source was constructed using the same model as described in Section \ref{sub:high_snr}, but we now set the peak intensities of the components of the source to $\bar{I}_0=4$ and $\bar{I}_1=\bar{I}_2=20$. These peak intensity levels were selected to provide $\sim$1000 pixels that met the signal-to-noise cut, which is consistent with the number of pixels in the cuts on the VLA data, ignoring the effects of the beam. The astrophysical scatter between the polarization angle, $\alpha$, and the total intensity gradient position angle, $\beta$, was drawn randomly from the distribution given in equation (\ref{eq:wc}), with a mean of zero and a dispersion of $\eta=40^{\circ}$.

 In anticipation of using real data, we measured the gradient of the total intensity distribution using a method of finite differencing:
\begin{align}\label{eq:grad_finite}
&\bm{\nabla}I_{\mathrm{obs}}\left(x_i,y_j\right)=\nonumber\\
&\frac{1}{2}\left[I\left(x_{i+1},y_j\right)-I\left(x_{i-1},y_j\right), I\left(x_i,y_{j+1}\right)-I\left(x_i,y_{j-1}\right)\right],
\end{align}
where the subscripts on $x$ and $y$ denote the pixel index. 

We recovered the estimated shear from $10^{4}$ noise realizations of the simulations discussed above. The results of this test are shown in the three middle panels of Figure \ref{fig:all_simple_sims}. The mean estimates were $\left<\hat{\gamma}_1\right>=\left(0.995\pm0.020\right)\times10^{-2}$, $\left<\hat{\gamma}_2\right>=\left(-2.14\pm0.02\right)\times10^{-2}$, $\left<\hat{\omega}\right>=-0.551^{\circ}\pm0.008^{\circ}$.

Here we see that there is clear evidence for a bias in $\hat{\gamma}_2$ and $\hat{\omega}$, with the bias being largest for the rotation estimate. This bias is a result of the measurement errors on the position angles, the contribution of an astrophysical scatter, and the finite differencing method used to measure the gradient. To see how these effects contribute a bias to the estimates, we can write the measured position angles in terms of the true position angles and an error, which may have a contribution from both a random  component, due to noise, and a systematic offset; in this case, a systematic offset is caused by the finite differencing method, but it could also have a contribution from beam convolution, as is the case when we consider real data in Section \ref{sec:results}. The observed position angles are then
\begin{align}\label{eq:angle_errs}
\hat{\alpha}_{\mathrm{obs}}=&\alpha_{\mathrm{obs}}+\delta\alpha_{\mathrm{obs}},\nonumber\\
\hat{\beta}_{\mathrm{obs}}=&\beta_{\mathrm{obs}}+\delta\beta_{\mathrm{obs}}.
\end{align}
For convenience we will assume that the true intrinsic linear polarization vector is aligned with the intrinsic gradient of $I$, so that the error term $\delta\alpha_{\mathrm{obs}}$ in equation (\ref{eq:angle_errs}) contains both the contribution from measurement errors on $Q$ and $U$, and the astrophysical scatter. The error term $\delta\beta_{\mathrm{obs}}$ has a contribution from measurement errors only. These errors on the angles propagate a bias into the mean sine and cosine terms in equations (\ref{eq:mat_B}) and (\ref{eq:vec_d}). To see why this is the case, we can follow the approach outlined in \cite{whittaker14} and write the expectation values of the cosine and sine of $\alpha_{\mathrm{obs}}$ as
\begin{equation}\label{eq:means_covs}
\left(
\begin{array}{c}
\left<\cos2\alpha_{\mathrm{obs}}\right> \\
\left<\sin2\alpha_{\mathrm{obs}}\right>
\end{array} \right)=
\frac{1}{\xi_{C}^2+\xi_{S}^2}\left(\begin{array}{c}
C'\xi_{C}+S'\xi_{S} \\
S'\xi_{C}-C'\xi_{S}
\end{array} \right),
\end{equation}
where
\begin{align}\label{eq:means_covs_def}
C'=&\left<\cos2\hat{\alpha}_{\mathrm{obs}}\right>-\mathrm{cov}\left(\cos2\alpha_{\mathrm{obs}},\cos2\delta\alpha_{\mathrm{obs}}\right)\nonumber\\
&+\mathrm{cov}\left(\sin2\alpha_{\mathrm{obs}},\sin2\delta\alpha_{\mathrm{obs}}\right),\nonumber\\
S'=&\left<\sin2\hat{\alpha}_{\mathrm{obs}}\right>-\mathrm{cov}\left(\sin2\alpha_{\mathrm{obs}},\cos2\delta\alpha_{\mathrm{obs}}\right)\nonumber\\
&-\mathrm{cov}\left(\cos2\alpha_{\mathrm{obs}},\sin2\delta\alpha_{\mathrm{obs}}\right),\nonumber\\
\xi_{C}=&\left<\cos2\delta\alpha_{\mathrm{obs}}\right>,\nonumber\\
\xi_{S}=&\left<\sin2\delta\alpha_{\mathrm{obs}}\right>.
\end{align}
A similar analysis can be applied to the means of all the cosine and sine terms in $\mathbf{B}$ and $\bm{d}$ of equations (\ref{eq:mat_B}) and (\ref{eq:vec_d}). From equations (\ref{eq:means_covs}) and (\ref{eq:means_covs_def}), we see that the means of the trigonometric functions are biased by terms that depend on the distributions of errors on the angles, and on the covariances between the errors and the true observed position angles (i.e. the observed position angles assuming zero errors and no astrophysical scatter between the polarization and the gradient of the total intensity). As the bias terms depend on $\alpha_{\mathrm{obs}}$, the level of bias will depend on the shape of the source. As we are interested in only the high signal-to-noise regions of the galaxies, we propose a method whereby we use the observed source as a template for a set of calibration simulations that provide estimates of the $\xi$ and covariance terms corresponding to each of the means in $\mathbf{B}$ and $\bm{d}$. Here we present the steps followed when creating our calibration simulations:

\begin{enumerate}
  \item Create an image of $P$ using the $Q$ and $U$ images of the lensed source.
  \item Measure the gradient of the total intensity image using finite differencing, and set the polarization position angles, $\alpha_{\mathrm{obs}}^{(i)}$, to be equal to the recovered gradient position angles, $\beta_{\mathrm{obs}}^{(i)}$, thereby enforcing the assumption that the shear and rotation signals are zero in the calibration simulations.
  \item Reconstruct the $I$, $Q$ and $U$ maps by adding noise at the same level as used in the original images; this includes the contribution from astrophysical scatter.
  \item Create a suite of Monte-Carlo noise realizations and estimate all of the required $\xi$ and covariance terms by applying the same signal-to-noise cuts to the calibration simulations and measuring the gradient using finite differencing.
\end{enumerate}

For these tests, the calibration simulations were created from the simulated lensed maps prior to the addition of background noise. This was done to reduce the number of calibration simulations required to test for biases at this level. For calibration, we require that the overall structure of the simulated source resembles the intrinsic structure of the source in order to recover an estimate of the $\xi$ and covariance terms in equation (\ref{eq:means_covs}) (and equivalently for the terms corresponding to the other mean cosine and sine terms in $\mathbf{B}$ and $\bm{d}$). As we are only interested in the high signal-to-noise regions of the source, we expect the contribution from background noise in the maps to the estimates of the $\xi$ and the covariance terms to be small and that the dominant contribution will be from astrophysical scatter. 
    
Using the approach described above, we recovered estimates of the $\xi$ and covariance terms from $10^5$ independent Monte-Carlo simulations. We then reanalyzed the simulations used to produce the middle panels in Figure \ref{fig:all_simple_sims}, but we calculated the mean trigonometric functions using equation (\ref{eq:means_covs}). To clarify, the measured values of the position angles, $\hat{\alpha}_{\mathrm{obs}}$ and $\hat{\beta}_{\mathrm{obs}}$, in equation (\ref{eq:means_covs}) were recovered from the same simulations used in the middle panels of Figure \ref{fig:all_simple_sims}, but the $\xi$ and covariance terms were estimated from the $10^5$ independent Monte-Carlo simulations under the assumption of zero input shear and rotation signals, as described above. The results of this test are shown in the bottom panels of  Figure \ref{fig:all_simple_sims}. The numerical results of this test were $\left<\hat{\gamma}_1\right>=\left(0.987\pm0.054\right)\times10^{-2}$, $\left<\hat{\gamma}_2\right>=\left(-2.00\pm0.05\right)\times10^{-2}$, $\left<\hat{\omega}\right>=-1.02^{\circ}\pm0.02^{\circ}$. Hence, we see that the bias in the estimates has been greatly reduced and is now consistent with zero. However, the errors on the shear estimates have increased by a factor of 2.65, and the errors on the rotation estimates have increased by a factor of 1.86. It should be noted that it is likely that equation (\ref{eq:grad_finite}) is a sub-optimal method for measuring $\beta$ and a more accurate method may increase the precision of the estimator.

To gain some insight into the nature of the errors on estimates of the shear and rotation, we can make progress by assuming that the source is approximately circular. For a large number of reliable (i.e. meeting the signal-to-noise requirements) pixels, the matrix $\mathbf{B}$ can then be approximated as
\begin{equation}\label{eq:mat_B_approx}
\mathbf{B}\approx\xi_{\alpha}\xi_{\beta}\left(
\begin{array}{ccc}
2 & 0 & 0 \\
0 & 1 & 0 \\
0 & 0 & 1 
\end{array} \right),
\end{equation}
where
\begin{align}
\xi_{\alpha}=&\left<\cos2\delta\alpha_{\mathrm{obs}}\right>,\nonumber\\
\xi_{\beta}=&\left<\cos2\delta\beta_{\mathrm{obs}}\right>.
\end{align}
The estimator is then
\begin{align}\label{eq:approx_est}
\hat{\omega}=&-\frac{1}{2\xi_{\alpha}\xi_{\beta}N}\sum_{k=1}^N\sin2\left(\hat{\alpha}_{\mathrm{obs}}^{(k)}-\hat{\beta}_{\mathrm{obs}}^{(k)}\right),\nonumber\\
\hat{\gamma}_1=&\frac{1}{\xi_{\alpha}\xi_{\beta}N}\sum_{k=1}^N\left[\xi_{\beta}\cos2\hat{\alpha}_{\mathrm{obs}}^{(k)}-\xi_{\alpha}\cos2\hat{\beta}_{\mathrm{obs}}^{(k)}\right],\nonumber\\
\hat{\gamma}_2=&\frac{1}{\xi_{\alpha}\xi_{\beta}N}\sum_{k=1}^N\left[\xi_{\beta}\sin2\hat{\alpha}_{\mathrm{obs}}^{(k)}-\xi_{\alpha}\sin2\hat{\beta}_{\mathrm{obs}}^{(k)}\right].
\end{align}
Using equations (\ref{eq:assumption}) and (\ref{eq:pre-exp}), we can write the trigonometric functions of $\alpha_{\mathrm{obs}}\left(\bm{x}\right)$ and $\beta_{\mathrm{obs}}\left(\bm{x}\right)$ in terms of $\beta\left(\bm{x}'\right)$ to first order in the shear and rotation as
\begin{align}\label{eq:cos_sin}
\cos2\alpha_{\mathrm{obs}}\left(\bm{x}\right)=&\cos2\beta\left(\bm{x}'\right)-\gamma_1-\gamma_1\cos4\beta\left(\bm{x}'\right)-\gamma_2\sin4\beta\left(\bm{x}'\right),\nonumber\\
\sin2\alpha_{\mathrm{obs}}\left(\bm{x}\right)=&\sin2\beta\left(\bm{x}'\right)-\gamma_2-\gamma_1\sin4\beta\left(\bm{x}'\right)+\gamma_2\cos4\beta\left(\bm{x}'\right),\nonumber\\
\cos2\beta_{\mathrm{obs}}\left(\bm{x}\right)=&\cos2\beta\left(\bm{x}'\right)-2\gamma_1-2\omega\sin2\beta\left(\bm{x}'\right),\nonumber\\
\sin2\beta_{\mathrm{obs}}\left(\bm{x}\right)=&\sin2\beta\left(\bm{x}'\right)-2\gamma_2+2\omega\sin2\beta\left(\bm{x}'\right).
\end{align}
Substituting equation (\ref{eq:cos_sin}) into equation (\ref{eq:approx_est}) and assuming a circular source, so that $\left<\cos2\beta\left(\bm{x}'\right)\right>=\left<\sin2\beta\left(\bm{x}'\right)\right>=0$, it can be shown that $<\hat{\bm{\Gamma}}>=\bm{\Gamma}$. Furthermore, as the estimator in equation (\ref{eq:approx_est}) is linear in the trigonometric functions, if the shear and rotation vary across the source, the recovered estimates are the means of the variable signals.

The linearity of the estimator in equation (\ref{eq:approx_est}) in terms of the mean cosine and sine terms also implies that the errors on the estimates will be approximately Gaussian distributed with the dispersions
\begin{align}\label{eq:errors_circ}
\sigma_{\hat{\omega}}^2=&\frac{1}{8\xi_{\alpha}^2\xi_{\beta}^2N}\left(1-\xi_{4\mathrm{tot}}\right),\nonumber\\
\sigma_{\hat{\gamma}_1}^2=&\sigma_{\hat{\gamma}_2}^2=\frac{1}{2\xi_{\alpha}^2\xi_{\beta}^2N}\left(\xi_{\alpha}^2+\xi_{\beta}^2-2\xi_{\alpha}^2\xi_{\beta}^2\right),
\end{align}
where 
\begin{equation}
\xi_{4\mathrm{tot}}=\left<\cos4\delta\alpha_{\mathrm{obs}}\right>\left<\cos4\delta\beta_{\mathrm{obs}}\right>.
\end{equation}

We see that the correction terms, $\xi$, correct for biasing due to the measurement errors by boosting the means of the trigonometric functions. This boosting effect propagates into the errors on the shear and rotation estimates, increasing the errors as compared with the uncorrected estimators. If one assumes that the difference between the angles $\alpha$ and $\beta$ is dominated by an astrophysical scatter component with a distribution described by equation (\ref{eq:wc}), then it can be shown that
\begin{equation}\label{eq:wc_beta}
\xi_{\alpha}=e^{-2\eta^2},\quad\xi_{\beta}=1,\quad\xi_{4\mathrm{tot}}=e^{-4\eta^2}.
\end{equation}
Hence, if we further assume that $\eta^2\ll1$, then equation (\ref{eq:errors_circ}) simplifies to
\begin{equation}\label{eq:errors_circ_simp}
\sigma_{\hat{\omega}}^2=\frac{\eta^2}{2N},\quad\sigma_{\hat{\gamma}_1}^2=\sigma_{\hat{\gamma}_2}^2=\frac{2\eta^2}{N}.
\end{equation}
The error on estimates of the shear due to galaxy shape noise when using the standard approach of averaging over galaxy shapes is $\sigma_{\gamma}^2=\sigma_{\epsilon}^2/N_{\mathrm{gal}}$, where $\sigma_{\epsilon}$ is the dispersion in intrinsic galaxy shapes and $N_{\mathrm{gal}}$ is the number of source galaxies. Hence, we see that the contribution from astrophysical scatter to the error on the shear estimates using our approach ($\eta$ in equation (\ref{eq:errors_circ_simp})) plays the same role as the intrinsic shape dispersion in the standard shape based approach.

In the bottom panels of Figure \ref{fig:all_simple_sims}, the red curves show Gaussian distributions using the means from the simulations and the dispersions given in equation (\ref{eq:errors_circ}). These plots show that, for the source model used, the distribution of errors on the estimates are well described by the Gaussian assumption and the dispersions predicted by equation (\ref{eq:errors_circ}).

The dominant contribution to the errors is expected to come from the astrophysical scatter as we only consider the high signal-to-noise regions of the sources. It is expected that the magnitude of the astrophysical scatter will have some dependence on the resolution at which the objects are observed. Hence, it may be possible to identify an optimal scale on which to perform this analysis. If the original image is made at a high resolution, one is free to smooth the image to the desired resolution if this smoothing step is also applied consistently to the calibration simulations when calculating the bias correction terms.

Equation (\ref{eq:means_covs}) was derived assuming an exact knowledge of the bias correction terms. In general these terms have to be estimated using both the data and the calibration simulations, and hence there will also be an error on the estimates of these terms.

In order to gain some insight into the impact of errors on the bias correction terms, let us assume that the distribution of $\alpha_{\mathrm{obs}}-\beta_{\mathrm{obs}}$ is dominated by an astrophysical scatter described by equation (\ref{eq:wc}). In this case, $\xi_{\beta}$ is given by equation (\ref{eq:wc_beta}). If we attempt to fit equation (\ref{eq:wc}) to the data, we will have an error on our estimate of $\eta^2$ due to the finite number of high signal-to-noise pixels used in the analysis. If we now assume that errors on $\eta^2$ are much less than $\eta^2$ and are drawn randomly from a Gaussian distribution with zero mean and variance $\sigma_{\eta}^2$, then errors on estimates of $\xi_{\beta}$ will follow a lognormal distribution. Ignoring any systematic effects associated with the details of the simulations, the estimator will be biased due to the error on $\eta^2$, so that
\begin{equation}
\hat{\bm{\Gamma}}=\bm{\Gamma}e^{2\sigma_{\eta}^2}.
\end{equation}
For the analysis in Section \ref{sec:results}, the contribution from this bias is expected to be well below the percent level. Errors on the estimates will also increase by a factor of $1+e^{4\sigma_{\eta}^2}$, which again leads to a fractional difference below the percent level for the analysis in Section \ref{sec:results}. Therefore, for the remainder of this paper, we will ignore contributions from errors on estimates of the bias correction terms.

\section{Characterization of the data}
\label{sec:character_data}
Here we describe the data used in the analysis performed in the following section, where we apply our estimator to observations made with the VLA and MERLIN array. The specific sources we use have FRII morphology, but this is not necessary and, as already pointed out, the technique can be applied to any well resolved source. We use these sources since the data is available. They were taken for other reasons, and hence are not necessarily optimal for this purpose. 

\begin{table*}
\begin{minipage}{6in}
\centering
\begin{tabular}{|c|c|c|c|c|c|c|}
\hline
Source & Resolution (arcsec) & Pixel width (arcsec) & RA (B1950) & Dec (B1950) & Redshift & Frequency (GHz)\\ [0.1ex]
\hline
3C6.1 & 0.25 & 0.065 & 0 13 34.4 & +79 0 11.1 & 0.84 & 8.46 \\ [0.1ex]
3C15 & 2.3 & 0.67 & 0 34 30.6 & -1 25 40.4 & 0.073 & 8.46 \\ [0.1ex]
3C20 & 0.22 & 0.050 & 0 40 20.1 & +51 47 10.2 & 0.17 & 8.41 \\ [0.1ex]
3C22 & 0.25 & 0.050 & 0 48 4.71 & +50 55 45.4 & 0.94 & 8.46 \\ [0.1ex]
3C34 & 0.40 & 0.070 & 1 7 32.5 & +31 31 23.9 & 0.69 & 4.87 \\ [0.1ex]
3C41 & 0.20 & 0.050 & 1 23 54.7 & +32 57 38.3 & 0.79 & 8.46 \\ [0.1ex]
3C47 & 1.3 & 0.20 & 1 33 40.4 & +20 42 10.2 & 0.43 & 4.89 \\ [0.1ex]
3C55 & 0.40 & 0.10 & 1 54 19.5 & +28 37 4.80 & 0.74 & 4.85 \\ [0.1ex]
3C67 & 0.050 & 0.015 & 2 21 18.0 & +27 36 37.2 & 0.31 & 4.99 \\ [0.1ex]
3C105 & 0.25 & 0.050 & 4 4 47.8 & +3 32 49.7 & 0.089 & 8.41 \\ [0.1ex]
3C111.1.6 & 1.6 & 0.20 & 4 15 1.10 & +37 54 37.0 & 0.049 & 8.35 \\ [0.1ex]
3C123 & 0.23 & 0.050 & 4 33 55.2 & +29 34 12.6 & 0.22 & 8.44 \\ [0.1ex]
3C132 & 0.220 & 0.050 & 4 53 42.2 & +22 44 43.9 & 0.21 & 8.44 \\ [0.1ex]
3C153 & 0.26 & 0.050 & 6 5 44.4 & +48 4 48.8 & 0.28 & 8.41 \\ [0.1ex]
3C175 & 0.25 & 0.060 & 7 10 15.4 & +11 51 24.0 & 0.77 & 8.45 \\ [0.1ex]
3C184 & 0.35 & 0.070 & 7 33 59.0 & +70 30 1.10 & 0.99 & 4.86 \\ [0.1ex]
3C184.1 & 2.5 & 0.80 & 7 34 25.0 & +80 33 24.1 & 0.12 & 8.47 \\ [0.1ex]
3C192 & 0.80 & 0.25 & 8 2 35.5 & +24 18 26.4 & 0.060 & 8.23 \\ [0.1ex]
3C196 & 0.35 & 0.075 & 8 9 59.4 & +48 22 7.57 & 0.87 & 4.86 \\ [0.1ex]
3C197.1 & 0.25 & 0.080 & 8 18 1.06 & +47 12 8.30 & 0.13 & 8.46 \\ [0.1ex]
3C207 & 0.35 & 0.060 & 8 38 1.72 & +13 23 5.57 & 0.68 & 4.86 \\ [0.1ex]
3C215 & 0.37 & 0.10 & 9 3 44.1 & +16 58 16.1 & 0.41 & 4.89 \\ [0.1ex]
3C216 & 0.25 & 0.065 & 9 6 17.3 & +43 5 58.6 & 0.67 & 8.21 \\ [0.1ex]
3C217 & 0.35 & 0.060 & 9 5 41.4 & +38 0 29.9 & 0.90 & 4.86 \\ [0.1ex]
3C220.1 & 0.25 & 0.050 & 9 26 31.9 & +79 19 45.4 & 0.61 & 8.44 \\ [0.1ex]
3C220.3 & 0.35 & 0.060 & 9 31 10.5 & +83 28 55.0 & 0.69 & 4.86 \\ [0.1ex]
3C223 & 0.25 & 0.080 & 9 38 18.0 & +39 58 20.2 & 0.11 & 8.47 \\ [0.1ex]
3C225B & 0.050 & 0.015 & 9 39 32.2 & +13 59 33.3 & 0.58 & 4.99 \\ [0.1ex]
3C226 & 0.20 & 0.050 & 9 41 36.2 & +10 0 3.80 & 0.82 & 8.46 \\ [0.1ex]
3C227 & 2.5 & 0.70 & 9 45 7.80 & +7 39 9.00 & 0.086 & 8.47 \\ [0.1ex]
3C234 & 0.30 & 0.060 & 9 58 57.4 & +29 1 37.4 & 0.19 & 8.44 \\ [0.1ex]
3C247 & 0.35 & 0.060 & 10 56 8.38 & +43 17 30.6 & 0.75 & 4.86 \\ [0.1ex]
3C249.1 & 0.35 & 0.10 & 11 0 27.3 & +77 15 8.62 & 0.31 & 4.89 \\ [0.1ex]
3C254 & 0.35 & 0.070 & 11 11 53.3 & +40 53 41.5 & 0.73 & 4.89 \\ [0.1ex]
3C263 & 0.35 & 0.065 & 11 37 9.30 & +66 4 27.0 & 0.66 & 4.86 \\ [0.1ex]
3C263.1 & 0.35 & 0.070 & 11 40 49.2 & +22 23 34.9 & 0.82 & 4.89 \\ [0.1ex]
3C265 & 0.40 & 0.070 & 11 42 52.4 & +31 50 29.1 & 0.81 & 4.85 \\ [0.1ex]
3C277.2 & 0.40 & 0.070 & 12 51 4.20 & +15 58 51.2 & 0.77 & 4.86 \\ [0.1ex]
3C280 & 0.35 & 0.070 & 12 54 41.7 & +47 36 32.7 & 1.0 & 4.89 \\ [0.1ex]
3C284 & 0.90 & 0.20 & 13 8 41.3 & +27 44 2.60 & 0.24 & 8.06 \\ [0.1ex]
3C289 & 0.35 & 0.070 & 13 43 27.4 & +50 1 32.0 & 0.97 & 4.89 \\ [0.1ex]
3C299 & 0.41 & 0.050 & 14 19 6.29 & +41 58 30.2 & 0.37 & 4.86 \\ [0.1ex]
3C319 & 0.90 & 0.20 & 15 22 43.9 & +54 38 38.4 & 0.19 & 8.44 \\ [0.1ex]
3C325 & 0.35 & 0.070 & 15 49 14.0 & +62 50 20.0 & 0.86 & 4.86 \\ [0.1ex]
3C336 & 0.35 & 0.060 & 16 22 32.2 & +23 52 2.00 & 0.93 & 4.86 \\ [0.1ex]
3C340 & 0.40 & 0.060 & 16 27 29.4 & +23 26 42.6 & 0.78 & 4.86 \\ [0.1ex]
3C349 & 2.9 & 0.80 & 16 58 4.44 & +47 7 20.3 & 0.21 & 8.44 \\ [0.1ex]
3C352 & 0.35 & 0.080 & 17 9 18.0 & +46 5 6.00 & 0.81 & 4.71 \\ [0.1ex]
3C381 & 0.25 & 0.060 & 18 32 24.5 & +47 24 39.0 & 0.16 & 8.44 \\ [0.1ex]
3C401 & 0.27 & 0.050 & 19 39 38.8 & +60 34 33.5 & 0.20 & 8.44 \\ [0.1ex]
3C403 & 0.75 & 0.20 & 19 49 44.1 & +2 22 41.5 & 0.059 & 8.47 \\ [0.1ex]
3C433 & 0.25 & 0.080 & 21 21 30.5 & +24 51 33.0 & 0.10 & 8.47 \\ [0.1ex]
3C438 & 0.23 & 0.050 & 21 53 45.5 & +37 46 12.8 & 0.29 & 8.44 \\ [0.1ex]
3C441 & 0.35 & 0.065 & 22 3 49.3 & +29 14 43.8 & 0.71 & 4.86 \\ [0.1ex]
3C452 & 0.25 & 0.080 & 22 43 32.8 & +39 25 27.3 & 0.081 & 8.46 \\ [0.1ex]
3C455 & 0.40 & 0.050 & 22 52 34.5 & +12 57 33.5 & 0.54 & 4.86 \\ [0.1ex]
4C14.11 & 0.23 & 0.070 & 4 11 40.9 & +14 8 48.3 & 0.21 & 8.44 \\ [0.1ex]
4C74.16 & 0.30 & 0.065 & 10 9 49.7 & +74 52 29.8 & 0.81 & 8.47 \\ [0.1ex]
 \hline
\end{tabular}
\caption{Physical properties of the FRII sample used in this paper. Details of where the data were obtained from are given in the main text.}
\label{table:full}
\end{minipage}
\end{table*}

The data consists of polarization and total intensity maps for 58 sources with redshifts in the range $0<z<1$. The majority of the data was retrieved from the \cite{mullin08} database\footnote{\href{zl1.extragalactic.info}{zl1.extragalactic.info}}, with the rest of the sources from data presented in \cite{leahy97} and \cite{black92}. The important physical properties of the full sample considered are given in Table \ref{table:full}. Where possible, the positions of the objects were obtained from the \cite{mullin08} database, and the remainder were found in \cite{tabara80}.

\begin{figure*}
\begin{minipage}{6in}
\includegraphics{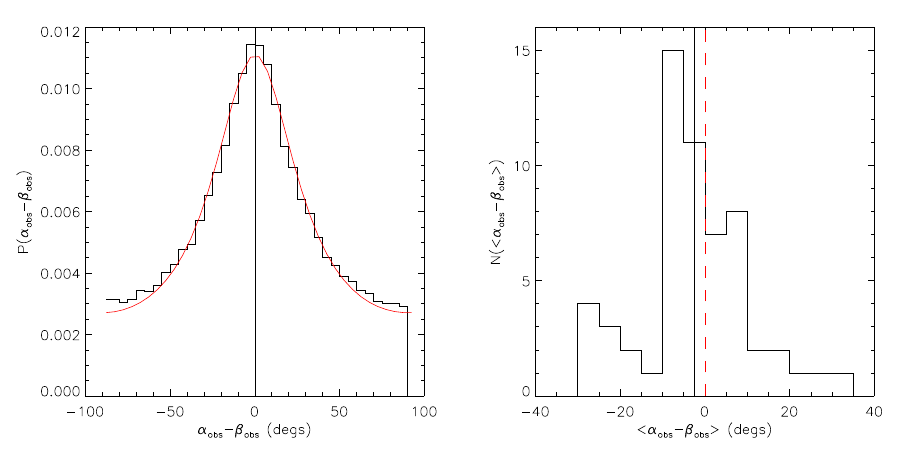}
\caption{\emph{Left-panel}: The distribution of $\alpha_{\mathrm{obs}}-\beta_{\mathrm{obs}}$ from the high signal-to-noise regions of the 58 FRII sources. For each source, the mean value of $\alpha_{\mathrm{obs}}-\beta_{\mathrm{obs}}$ has been subtracted from each value of $\alpha$ to remove any systematic astrophysical effects; {this step was carried out for presentation purposes only and was not included when applying the estimator to the data in Section \ref{sec:results}.} The total number of angles included in the histogram is 119742. The vertical black line indicates zero difference. The red curve is the distribution given in equation (\ref{eq:wc}), with best-fit dispersion parameter $\eta_{\mathrm{total}}=42.2^{\circ}$. \emph{Right-panel}: The distribution of the means of $\alpha_{\mathrm{obs}}-\beta_{\mathrm{obs}}$ for the 58 sources, which were removed from the left-hand figure. The vertical black line is the mean, weighted by the number of pixels used in each source. The mean value is $-2.08^{\circ}$. {Numerical values of the means of $\alpha_{\mathrm{obs}}-\beta_{\mathrm{obs}}$ for the sources included in the results discussed in Section \ref{sec:results} are given in table \ref{table:derived}.}}
\label{fig:full_relationship}
\end{minipage}
\end{figure*}

\begin{figure*}
\begin{minipage}{6in}
\includegraphics{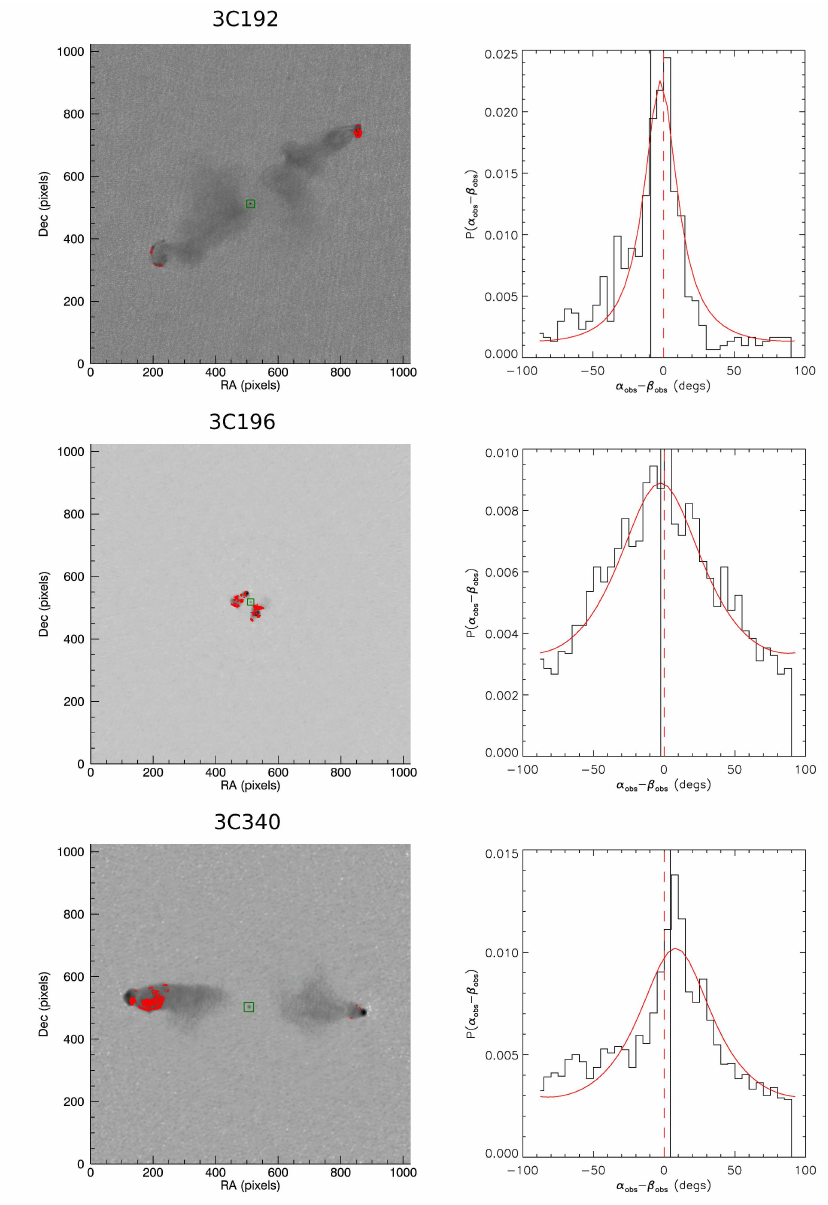}
\caption{The total intensity distributions for the three sources 3C192, 3C196, and 3C340 are shown in the left-hand panels. In red we show the pixels which satisfy the signal-to-noise cuts. The right-hand panels show the distributions of $\alpha_{\mathrm{obs}}-\beta_{\mathrm{obs}}$ in the red regions for each source. The red curves show equation (\ref{eq:wc}) with the dispersion parameters, $\eta_{\mathrm{total}}$, fitted to the data and given in Table \ref{table:derived}. The green squares show masked regions which contain the source AGNs.}
\label{fig:tot_intense}
\end{minipage}
\end{figure*}
For our analysis, we began by first inspecting the source images by eye and, where possible, masking the regions in the images containing the central source AGN; these regions have a high signal-to-noise, but they are unresolved. We also selected regions of the image where the radio emission from the source is expected to be negligible so that these regions could be used to measure the noise in the total intensity and polarization maps. We then measured the gradient of the total intensity images and selected the regions of the maps which satisfied the signal-to-noise requirements discussed in Section \ref{sub:high_snr}. The distribution of $\alpha_{\mathrm{obs}}-\beta_{\mathrm{obs}}$ from the 58 sources is shown in left-panel of Figure \ref{fig:full_relationship}. The red curve is equation (\ref{eq:wc}) fitted to the data, with best-fit parameter $\eta=42.2^{\circ}$. The right-panel is the distribution of recovered means of $\alpha_{\mathrm{obs}}-\beta_{\mathrm{obs}}$ for each source. {Numerical values of the means of $\alpha_{\mathrm{obs}}-\beta_{\mathrm{obs}}$ for the sources considered in Section \ref{sec:results} are given in table \ref{table:derived} along with estimates of the uncertainties.} Three examples of the total intensity images used are shown in Figure \ref{fig:tot_intense} along with the corresponding distribution of $\alpha_{\mathrm{obs}}-\beta_{\mathrm{obs}}$. These three examples were selected as they demonstrate a range of different dispersions between $\alpha_{\mathrm{obs}}-\beta_{\mathrm{obs}}$.

In order to estimate the contribution from astrophysical scatter, we adopted a method of forward modeling using simulations constructed from the data via the approach discussed in Section \ref{sub:high_snr}. We produced a set of simulated $I$, $Q$ and $U$ maps assuming Gaussian noise, with the noise levels set to match the levels measured from the images. An assumed level of astrophysical scatter was then added to the polarization position angle using the distribution given in equation (\ref{eq:wc}) with a constant dispersion, $\eta_{\mathrm{astro}}$. The simulated maps were smoothed with a Gaussian beam, and the width of the beam matched the width of the {\tt CLEAN} beam applied when creating the images. The convolution of the noisy simulated map with a Gaussian beam produces noise correlations between neighboring pixels. The detail of noise in radio images is more complicated than the noise in our calibration simulations, but as we are only interested in the high signal-to-noise regions, we expect our simulations to be sufficient.

The gradient of the total intensity images and the polarization position angles were recovered from the high signal-to-noise regions of these maps for $10^3$ noise realizations. A histogram was constructed using the $\alpha_{\mathrm{obs}}-\beta_{\mathrm{obs}}$ values recovered from the simulations and a binsize of $5^{\circ}$, and equation (\ref{eq:wc}) was fitted to this histogram providing a corresponding dispersion parameter for the total contribution of errors and astrophysical scatter, $\eta_{\mathrm{total}}$. This was repeated for a range of values of $\eta_{\mathrm{astro}}$. The estimated value of $\eta_{\mathrm{astro}}$ for each source was taken to be the one which provided the value of $\eta_{\mathrm{total}}$ in the fit to the histogram of simulated $\alpha_{\mathrm{obs}}-\beta_{\mathrm{obs}}$ that matched that of the true maps. This procedure was repeated for each source in table \ref{table:full}.

\begin{figure}
\includegraphics{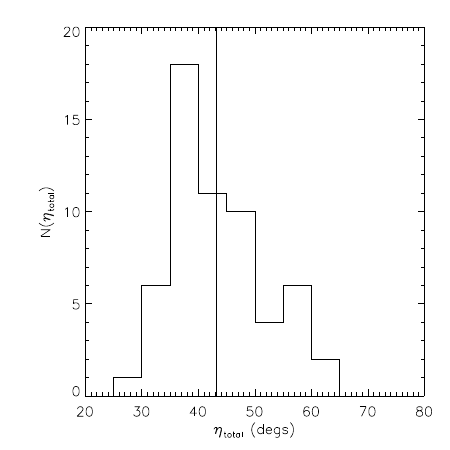}
\caption{The distribution of $\eta_{\mathrm{total}}$ fitted to histograms of the measured $\alpha_{\mathrm{obs}}-\beta_{\mathrm{obs}}$ from the image data for the 58 sources in the sample. The vertical line shows the mean value.}
\label{fig:sigmas}
\end{figure}

The distribution of $\eta_{\mathrm{total}}$ fitted to the 58 sources in the sample is shown in Figure \ref{fig:sigmas}. The mean value is $\left<\eta_{\mathrm{total}}\right>=43.1^{\circ}$. This is a weighted mean, with the weights taken to be
\begin{equation}\label{eq:neff_weights}
\mathrm{w}_i=N^{(i)}_{\mathrm{eff}}=\frac{2\ln2}{\pi}\frac{N_i}{\left(W_i/P_i\right)^2},
\end{equation}
where we define $N^{(i)}_{\mathrm{eff}}$ as the effective number of reliable pixels for source $i$. Here, $N_i$ is the number of pixels in the maps above the signal-to-noise threshold for the $i^{\mathrm{th}}$ source , $W_i$ is the beam width (FWHM), and $P_i$ is the pixel width. The ratio $W_i/P_i$ has been included as an estimate of the contribution of the beam to the number of effective pixels in the sample. The exact effect of the beam on the number of effective pixels is difficult to quantify when considering the errors on angle measurements, as the errors depend on the morphology of the Stokes parameter distributions. For this analysis we assume that the errors on the angles scale proportionally to the area of the beam, which, given a constant level of noise, describes how the beam effects the number of effective pixels when measuring $<$$Q$$>$ and $<$$U$$>$.

The effects of the beam (the effects of pixelization will be sub-dominant as the pixel width is less than the beam width) on the mean sines and cosines in equations (\ref{eq:mat_B}) and (\ref{eq:vec_d}) are captured by the $\xi$ and covariance terms in equation (\ref{eq:means_covs}) which we derive using calibration simulations, as discussed in Section \ref{sub:real_snr}. Therefore, to calibrate for both noise bias and the bias introduced by beam convolution, we created a suite of $10^4$ calibration simulations for each source using the approach discussed in Section \ref{sub:real_snr}. As discussed above, we assume a Gaussian error distribution for the noise on the $I$, $Q$ and $U$ maps, and we also assume a Gaussian beam when convolving the maps with the beam. The beam convolution step is applied directly after the noise is added.

Beam convolution has two undesirable effects when creating the calibration simulations. Firstly, it smooths the noise, and hence reduces the apparent values of $\sigma_{P}$ and $\sigma_{\mathrm{b}}$ in the convolved maps. To ameliorate this problem, we scaled the noise so that the noise applied to the calibration simulations is
\begin{align}
\sigma_{\mathrm{b}}^{\mathrm{calib}}=&\frac{\sigma_{\mathrm{b}}}{\sqrt{\sum_iK^2\left(\bm{x}_i\right)}},\nonumber\\
\sigma_{P}^{\mathrm{calib}}=&\frac{\sigma_{P}}{\sqrt{\sum_iK^2\left(\bm{x}_i\right)}},
\end{align}
where $K$ is the Kernel describing the shape of the beam. This scaling provides the correct level of noise in the beam convolved simulated maps. The other effect is that the beam also reduces the apparent intensity of the maps. This reduction in intensity means that regions of the calibration images which are included in the analysis are omitted from the calibration simulations as they do not meet the signal-to-noise requirements. This propagates as a bias into estimates of the calibration terms. For this analysis, we multiplied the simulated images by a scaling factor so that the peak intensities of the smoothed calibration simulations, before the addition of noise, matched the peak intensities of the true images. At some level, this step attempts to correct for the fact that the calibration images are constructed from images that have already been convolved with the beam.

Ideally, the calibration simulations would be constructed from the unconvolved images and include the correct shear, rotation and systematics. However, in reality this is obviously not possible. We have confirmed that calibrating the mean cosines and sines using equation (\ref{eq:means_covs}) completely corrects for biases due to beam convolution if the ideal scenario is assumed. However, there is some residual biasing if one assumes the more realistic approach described above due to the imperfect calibration simulations. This bias, however, was found to be much smaller than the errors on the shear and rotation in all of the tests performed.

From these calibration simulations, we estimate the $\xi$ and covariance terms in equation (\ref{eq:means_covs}), using finite differencing to measure the gradients and including the correct signal-to-noise cuts, and correct the mean trigonometric functions accordingly.

\section{Results of applying the estimator to the data}
\label{sec:results}
Here we discuss the results of applying the estimator, with the calibration steps described in the previous section, to the 58 objects shown in table \ref{table:full}. We present results for the estimated shear and rotation, and also estimate the shear and rotation two point correlation functions (2PCFs).

\begin{table*}
\begin{minipage}{6in}
\centering
\begin{tabular}{|c|c|c|c|c|c|c|c|}
\hline
Source number & Source & $\hat{\gamma}_1$ & $\hat{\gamma}_2$ & $\hat{\omega}\,(\mathrm{degs})$ & $\eta_{\mathrm{total}}\,(\mathrm{degs})$ & $\left<\alpha_{\mathrm{obs}}-\beta_{\mathrm{obs}}\right>$ & $N_{\mathrm{eff}}$ \\ [0.1ex]
\hline
1 & 3C6.1 & $0.098\pm0.163$ & $-0.266\pm0.148$ & $0.92\pm3.73$ & $34.0\pm0.8$ & $1.37\pm5.43$ & 39.1 \\ [0.1ex]
2 & 3C15 & $0.377\pm0.245$ & $-0.804\pm0.231$ & $-2.51\pm6.47$ & $38.0\pm0.9$ & $-5.34\pm5.45$ & 48.6 \\ [0.1ex]
3 & 3C20 & $-0.232\pm0.103$ & $-0.222\pm0.099$ & $-3.21\pm2.43$ & $37.6\pm0.5$ & $-1.87\pm3.44$ & 119.0 \\ [0.1ex]
4 & 3C34 & $-0.233\pm0.448$ & $-0.279\pm0.406$ & $-6.35\pm12.4$ & $44.9\pm0.9$ & $-3.80\pm7.95$ & 31.8 \\ [0.1ex]
5 & 3C41 & $0.188\pm0.291$ & $0.589\pm0.296$ & $3.55\pm6.48$ & $45.7\pm0.8$ & $-2.39\pm5.13$ & 79.3 \\ [0.1ex]
6 & 3C47 & $-0.0277\pm0.0803$ & $-0.126\pm0.095$ & $-8.94\pm2.09$ & $34.4\pm0.3$ & $-9.71\pm3.19$ & 116 \\ [0.1ex]
7 & 3C105 & $0.184\pm0.317$ & $-0.339\pm0.297$ & $-10.7\pm7.56$ & $44.5\pm0.7$ & $-5.80\pm5.24$ & 72.2 \\ [0.1ex]
8 & 3C132 & $0.236\pm0.169$ & $-0.638\pm0.168$ & $10.8\pm4.1$ & $39.5\pm0.7$ & $-2.29\pm5.14$ & 59.1 \\ [0.1ex]
9 & 3C153 & $-0.135\pm0.318$ & $0.288\pm0.310$ & $5.73\pm6.52$ & $43.5\pm0.7$ & $5.46\pm6.25$ & 48.5 \\ [0.1ex]
10 & 3C184.1 & $0.207\pm0.309$ & $-0.089\pm0.311$ & $-3.73\pm7.13$ & $41.9\pm1.5$ & $-6.93\pm7.77$ & 29.1 \\ [0.1ex]
11 & 3C192 & $0.222\pm0.133$ & $0.006\pm0.123$ & $-8.48\pm3.87$ & $29.1\pm1.1$ & $-9.27\pm5.68$ & 26.2 \\ [0.1ex]
12 & 3C196 & $-0.491\pm0.425$ & $0.080\pm0.403$ & $-10.4\pm8.9$ & $47.1\pm0.8$ & $-2.53\pm5.77$ & 66.5 \\ [0.1ex]
13 & 3C217 & $-0.380\pm0.596$ & $-0.096\pm0.514$ & $-13.8\pm14.0$ & $43.3\pm0.9$ & $-6.33\pm8.92$ & 23.6 \\ [0.1ex]
14 & 3C220.1 & $-0.205\pm0.235$ & $-0.345\pm0.244$ & $0.10\pm5.37$ & $42.3\pm0.7$ & $-7.75\pm5.81$ & 53.1 \\ [0.1ex]
15 & 3C223 & $0.307\pm0.110$ & $-0.118\pm0.108$ & $4.47\pm3.06$ & $32.5\pm0.9$ & $-2.61\pm4.73$ & 47.1 \\ [0.1ex]
16 & 3C226 & $-0.594\pm0.356$ & $-0.037\pm0.353$ & $-3.85\pm7.91$ & $43.8\pm1.0$ & $5.43\pm6.36$ & 47.6 \\ [0.1ex]
17 & 3C227 & $-0.013\pm0.237$ & $0.069\pm0.254$ & $5.14\pm5.78$ & $37.1\pm1.2$ & $6.73\pm7.05$ & 27.7 \\ [0.1ex]
18 & 3C234 & $0.081\pm0.150$ & $0.040\pm0.145$ & $3.83\pm3.39$ & $34.7\pm0.7$ & $4.17\pm5.65$ & 37.7 \\ [0.1ex]
19 & 3C249.1 & $-0.333\pm0.114$ & $0.249\pm0.118$ & $-10.9\pm2.9$ & $36.4\pm0.7$ & $-7.05\pm4.29$ & 71.8 \\ [0.1ex]
20 & 3C263 & $0.095\pm0.168$ & $-0.105\pm0.174$ & $7.90\pm3.75$ & $36.1\pm0.6$ & $7.88\pm5.77$ & 39.3 \\ [0.1ex]
21 & 3C265 & $-0.173\pm0.188$ & $0.151\pm0.186$ & $12.0\pm4.3$ & $37.9\pm0.5$ & $14.5\pm5.3$ & 52.1 \\ [0.1ex]
22 & 3C336 & $-0.014\pm0.130$ & $-0.124\pm0.131$ & $8.49\pm3.18$ & $35.0\pm0.4$ & $8.03\pm4.32$ & 65.6 \\ [0.1ex]
23 & 3C340 & $0.485\pm0.283$ & $0.292\pm0.274$ & $-3.89\pm5.94$ & $43.2\pm0.5$ & $4.49\pm6.06$ & 50.7 \\ [0.1ex]
24 & 3C349 & $0.131\pm0.109$ & $-0.289\pm0.106$ & $0.29\pm2.90$ & $31.8\pm0.7$ & $-2.25\pm4.39$ & 52.6 \\ [0.1ex]
25 & 3C352 & $0.231\pm0.285$ & $-0.031\pm0.277$ & $2.44\pm6.21$ & $39.8\pm0.9$ & $1.83\pm6.48$ & 37.7 \\ [0.1ex]
26 & 3C381 & $0.153\pm0.178$ & $0.662\pm0.207$ & $13.0\pm5.2$ & $36.2\pm0.8$ & $-1.04\pm5.66$ & 40.8 \\ [0.1ex]
27 & 3C401 & $0.009\pm0.159$ & $0.475\pm0.166$ & $-12.3\pm4.1$ & $38.2\pm0.6$ & $-5.21\pm5.97$ & 40.9 \\ [0.1ex]
28 & 3C403 & $-0.039\pm0.302$ & $0.384\pm0.288$ & $-3.93\pm5.94$ & $46.1\pm0.8$ & $2.16\pm4.80$ & 92.2 \\ [0.1ex]
29 & 3C433 & $0.253\pm0.175$ & $0.330\pm0.237$ & $-5.50\pm5.33$ & $42.8\pm0.9$ & $0.36\pm4.50$ & 90.3 \\ [0.1ex]
30 & 3C441 & $-0.071\pm0.125$ & $-0.470\pm0.124$ & $-18.6\pm3.1$ & $35.8\pm0.5$ & $-27.1\pm4.8$ & 55.8 \\ [0.1ex]
\hline
\end{tabular}
\caption{Derived properties from the FRII sample. The source number identifies the source with the estimates displayed in Figure \ref{fig:ests_cut}. The columns labeled $\hat{\gamma}_1$, $\hat{\gamma}_2$, and $\hat{\omega}$ show the estimates for the two components of the shear and the rotation respectively for each of the sources. The column labeled $\eta_{\mathrm{total}}$ shows the total dispersion parameters fitted to the distributions of $\alpha_{\mathrm{obs}}-\beta_{\mathrm{obs}}$, {$\left<\alpha_{\mathrm{obs}}-\beta_{\mathrm{obs}}\right>$ shows the means of the measured $\alpha_{\mathrm{obs}}-\beta_{\mathrm{obs}}$ for each source (the error bars are estimates as $\eta_{\mathrm{total}}/\sqrt{N_{\mathrm{eff}}}$),} and $N_{\mathrm{eff}}$ shows the number of effective pixels used in the analysis of each source.}
\label{table:derived}
\end{minipage}
\end{table*}

\begin{figure*}
\begin{minipage}{6in}
\includegraphics{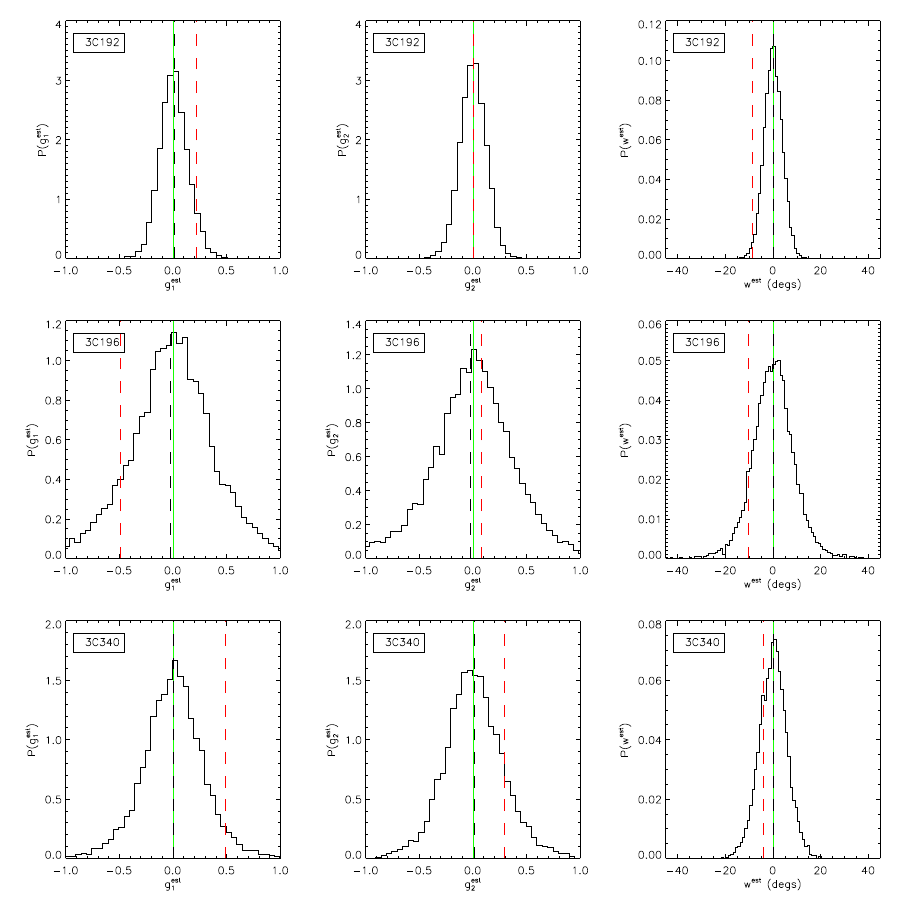}
\caption{The recovered shear and rotation estimates for each of the sources shown in Figure \ref{fig:tot_intense}. In each panel, the recovered estimate is shown as the red dashed line. The green line indicates a zero signal. The black histogram shows the recovered estimates from $10^4$ simulated noise realizations, as discussed in the main text. The black dashed line is the mean estimate from the simulations. The numerical results for all of the sources are given in table \ref{table:derived}.}
\label{fig:sim_dists}
\end{minipage}
\end{figure*}
We began by applying the estimator to each of the objects and recovering shear and rotation estimates. In order to quantify the errors on the estimates, we created simulated images using the method used to produce the calibration simulations, and assuming zero input shear and rotation signals. We then applied the estimator, including the bias corrections, to $10^4$ simulated noise realizations of each image and recovered the estimates. The recovered estimates from the real data and the simulations corresponding to the three sources shown in Figure \ref{fig:tot_intense} are presented in Figure \ref{fig:sim_dists}. 

For some of the objects, we find that there are substantial outliers in the estimates recovered from the simulations used to determine the errors. These were due to catastrophic failures in the simulation process. This leads to a calculated dispersion of the estimates which is not representative of the bulk of the distribution. In order to recover a more accurate estimate of the error, we chose to make cuts on the simulated distributions. First, we cut the distributions of the simulated shears and rotations so that we excluded any realizations where either component of the shear lies more than 3$\sigma$ from the corresponding mean value. We then compared the dispersions of the cut shear distributions with the original dispersions. If the difference between either of the two pairs of dispersions was greater than 0.05 of the original value (for an underlying Gaussian distribution, the corresponding difference would be $0.015\sigma$, we chose $0.05\sigma$ to allow for deviations from Gaussianity), the suite of simulations was replaced by the cut suite. This procedure was iterated until the cut distributions were consistent with the uncut distributions.

Two further cuts were made to create the final sample of shear and rotation estimates. The first was a cut on the sample so that we removed any source with $N_{\mathrm{eff}}<21$, where the definition of $N_{\mathrm{eff}}$ is given in equation (\ref{eq:neff_weights}). This cut was made as the shear and rotation estimates were deemed unreliable for low numbers of reliable pixels due to large dispersions and substantial outliers in the simulated distributions. The second cut was to remove any source for which $\eta_{\mathrm{total}}>0.7$. Large dispersions between $\alpha_{\mathrm{obs}}$ and $\beta_{\mathrm{obs}}$ due to measurement errors and astrophysical scatter produce large biases in the estimates. These biases, in turn, require large bias corrections which again produce large outliers in the shear estimates; therefore, the cut on $\eta_{\mathrm{total}}$ is made to avoid this problem. The cuts were made so that all shear and rotations deemed problematic were excluded from the sample. Ultimately such estimates would have been downweighted in subsequent use of the results, but it seemed prudent to seclude them from the beginning using well motivated cuts. For future surveys, where a larger number of sources will be available, a more objective method of cutting the sample must be devised.

\begin{figure}
\includegraphics{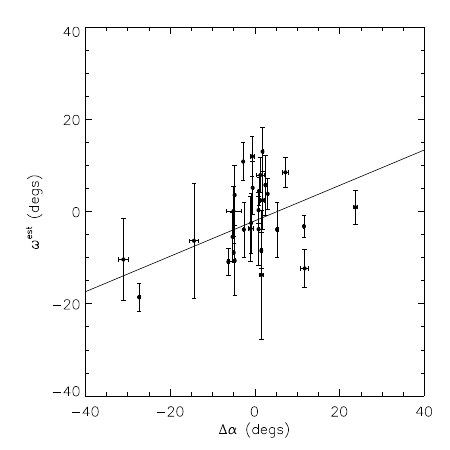}
\caption{{The recovered rotation estimate and corresponding $\Delta\alpha$ (defined in equation \ref{eq:Dalpha}) for each of the sources in table \ref{table:derived}. The line is the best-fit to the data. The Pearson's correlation coefficient was calculated to be $\rho_{\mathrm{corr}}=0.39\pm0.19$, where the errors have been estimated by constructing a distribution of $10^5$ correlation coefficients using random permutations of the set of rotation estimates with respect to the set of $\Delta\alpha$. This indicates that there is a positive correlation between $\Delta\alpha$ and the rotation estimate, as expected.}}
\label{fig:Fvom}
\end{figure}
The final list of derived quantities for the cut sample is given in table \ref{table:derived}. {The shear and rotation estimates in table \ref{table:derived} have been recovered under the assumption that the contribution from Faraday rotation is negligible. If one assumes that the observed Faraday rotation signal across a source is constant, one would expect the recovered rotation estimate for that source to be the sum of the Faraday rotation signal and the sought after cosmological rotation signal. In reality, however, the observed Faraday rotation signal varies across the sources, and our method will be sensitive to these variations. It should also be noted that systematic effects may arise due to the fact that we only use high signal-to-noise regions of the sources; however, we attempt to calibrate for these effects by mimicking the cuts in the calibration simulations.

Figure \ref{fig:Fvom} shows the estimated rotation signal for the 30 sources presented in table \ref{table:derived} plotted against $\Delta\alpha$, where we define
\begin{equation}\label{eq:Dalpha}
\Delta\alpha=\mathrm{RM}\times\lambda_{\mathrm{obs}}^2,
\end{equation}
and where RM is the best-fit rotation measure for an observed source \citep{simnor81} and $\lambda_{\mathrm{obs}}$ is the wavelength of the observation. Where possible, the rotation measures used in Figure \ref{fig:Fvom} were obtained from \cite{simnor81}, and the rest are from \cite{tabara80}. If we assume that the Faraday rotation across each source is constant and that there is zero contribution from other systematics and cosmological rotation, one would expect to see a one-to-one relationship between $\hat{\omega}$ and $\Delta\alpha$ in Figure \ref{fig:Fvom}; however, variations in Faraday rotation across each source at the pixel level reduces this correlation. Therefore, in order to remove contamination from Faraday rotation in future applications of our method, rotation measures should be mapped across each source and corrections for Faraday rotation made at the pixel level; uncertainties on these corrections must also be included at the pixel level in the calibration simulations.

In table \ref{table:derived}, we see that there are a number of sources which show a significant rotation detection. The source of this signal is expected to be dominated by Faraday rotation, and at present we do not have the power to discriminate between Faraday rotation and birefringence. The most striking example is the source 3C441, which shows a $\sim$6$\sigma$ detection of a rotation signal. However, for this source we see a large value of $<$$\alpha_{\mathrm{obs}}-\beta_{\mathrm{obs}}$$>$ which we also found to be approximately equal to $\Delta\alpha$, and so we expect that the estimated rotation signal for this source is dominated by Faraday rotation.}

\begin{figure*}
\begin{minipage}{6in}
\includegraphics{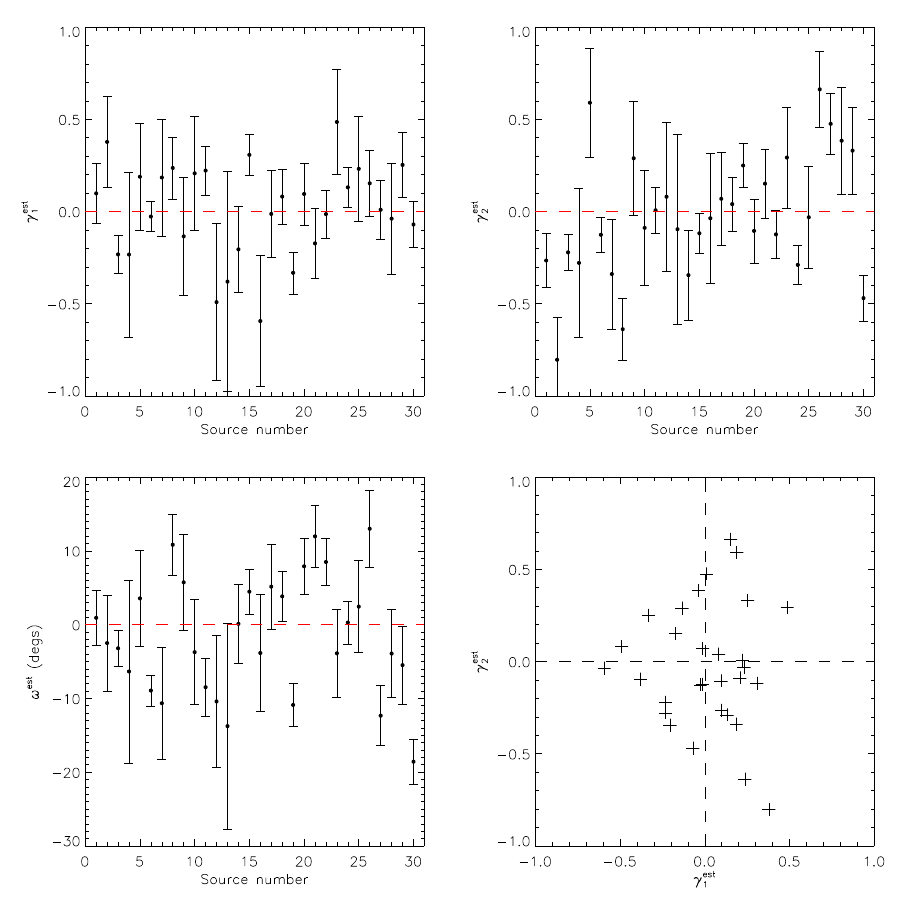}
\caption{The shear and rotation estimates for all of the sources in the final sample given in table \ref{table:derived}. The x-axis of the top and bottom-left panels indicates the corresponding source, with the identification numbers defined in the first column of table \ref{table:derived}. The bottom-right panel shows the shear estimates in the $\gamma_1-\gamma_2$ plane. From these plots we see no clear indication of residual systematics in the estimates.}
\label{fig:ests_cut}
\end{minipage}
\end{figure*}

The recovered shear and rotation estimates are displayed in Figure \ref{fig:ests_cut}. Using the inverse variance of each estimate as a weighting, the means of recovered shear and rotation estimates are \\$\left<\hat{\gamma}_1\right>=\left(2.27\pm3.00\right)\times10^{-2}$, $\left<\hat{\gamma}_2\right>=\left(-8.79\pm3.05\right)\times10^{-2}$, and $\left<\hat{\omega}\right>=-2.03^{\circ}\pm0.75^{\circ}$.

For this analysis we have assumed a perfect knowledge of the bias correction terms; however, as these terms are estimated using simulations based on the observed images, there will be additional uncertainties due to these terms. Assuming an accurate model of the beam and that the dispersion of $\alpha_{\mathrm{obs}}-\beta_{\mathrm{obs}}$ is dominated by astrophysical scatter, it is expected that errors in the bias correction terms will primarily come from errors in the astrophysical scatter model assumed which, in turn, is dominated by errors on the fit to the distribution of $\alpha_{\mathrm{obs}}-\beta_{\mathrm{obs}}$. Assuming that the distribution of $\alpha_{\mathrm{obs}}-\beta_{\mathrm{obs}}$ follows equation (\ref{eq:wc}), we can estimate the errors on the estimates of $\eta_{\mathrm{total}}$ by assuming Poisson errors on the values in each bin of the histogram of measured $\alpha_{\mathrm{obs}}-\beta_{\mathrm{obs}}$ to which the model is fitted and assuming that the fitted histogram is the true model. Using this approach, we estimated the errors on the fitted values of $\eta_{\mathrm{total}}$; these errors are shown in table \ref{table:derived}. We find that the fractional errors on $\eta_{\mathrm{total}}$ are less than 10\% in all cases.

Using the estimates presented in table \ref{table:derived} and the source positions from table \ref{table:full}, we calculated the two-point correlation functions (2PCFs) for the sample. The 2PCFs considered were
\begin{align}\label{eq:shear_2pf}
\xi_{\pm}\left(\theta\right)=&\frac{\sum\mathrm{w}_{\mathrm{a}}\mathrm{w}_{\mathrm{b}}\left[\gamma_{\mathrm{t}}\left(\bm{x}_{\mathrm{a}}\right)\gamma_{\mathrm{t}}\left(\bm{x}_{\mathrm{b}}\right)\pm\gamma_{\mathrm{x}}\left(\bm{x}_{\mathrm{a}}\right)\gamma_{\mathrm{x}}\left(\bm{x}_{\mathrm{b}}\right)\right]}{\sum\mathrm{w}_{\mathrm{a}}\mathrm{w}_{\mathrm{b}}},\nonumber\\
\xi_{\omega}\left(\theta\right)=&\frac{\sum\mathrm{w}^{\omega}_{\mathrm{a}}\mathrm{w}^{\omega}_{\mathrm{b}}\omega\left(\bm{x}_{\mathrm{a}}\right)\omega\left(\bm{x}_{\mathrm{b}}\right)}{\sum\mathrm{w}^{\omega}_{\mathrm{a}}\mathrm{w}^{\omega}_{\mathrm{b}}},\nonumber\\
\xi_{\gamma\omega}\left(\theta\right)=&\frac{\sum\left(\mathrm{w}_{\mathrm{a}}\mathrm{w}^{\omega}_{\mathrm{b}}+\mathrm{w}^{\omega}_{\mathrm{a}}\mathrm{w}_{\mathrm{b}}\right)\left(\left|\bm{\gamma}\right|\left(\bm{x}_{\mathrm{a}}\right)\omega\left(\bm{x}_{\mathrm{b}}\right)+\omega\left(\bm{x}_{\mathrm{a}}\right)\left|\bm{\gamma}\right|\left(\bm{x}_{\mathrm{b}}\right)\right)}{2\sum\left(\mathrm{w}_{\mathrm{a}}\mathrm{w}^{\omega}_{\mathrm{b}}+\mathrm{w}^{\omega}_{\mathrm{a}}\mathrm{w}_{\mathrm{b}}\right)},\nonumber\\
\end{align}
where the weights were taken to be the inverse mean square variance of the estimates. The tangential and cross shear $\gamma_{\mathrm{t}}$ and $\gamma_{\mathrm{x}}$ are the components of the shear rotated into a frame of reference which aligns with the separation vector $\bm{\theta}=\bm{x}_{\mathrm{a}}-\bm{x}_{\mathrm{b}}$. The modulus of the separation vector, $\theta$, was calculated for each source pair, and the source pairs were then binned according to their separation into bins of width $\Delta\theta=45^{\circ}$. The summations in equation (\ref{eq:shear_2pf}) were carried out over all source pairs for each bin.  

\begin{figure*}
\begin{minipage}{6in}
\includegraphics{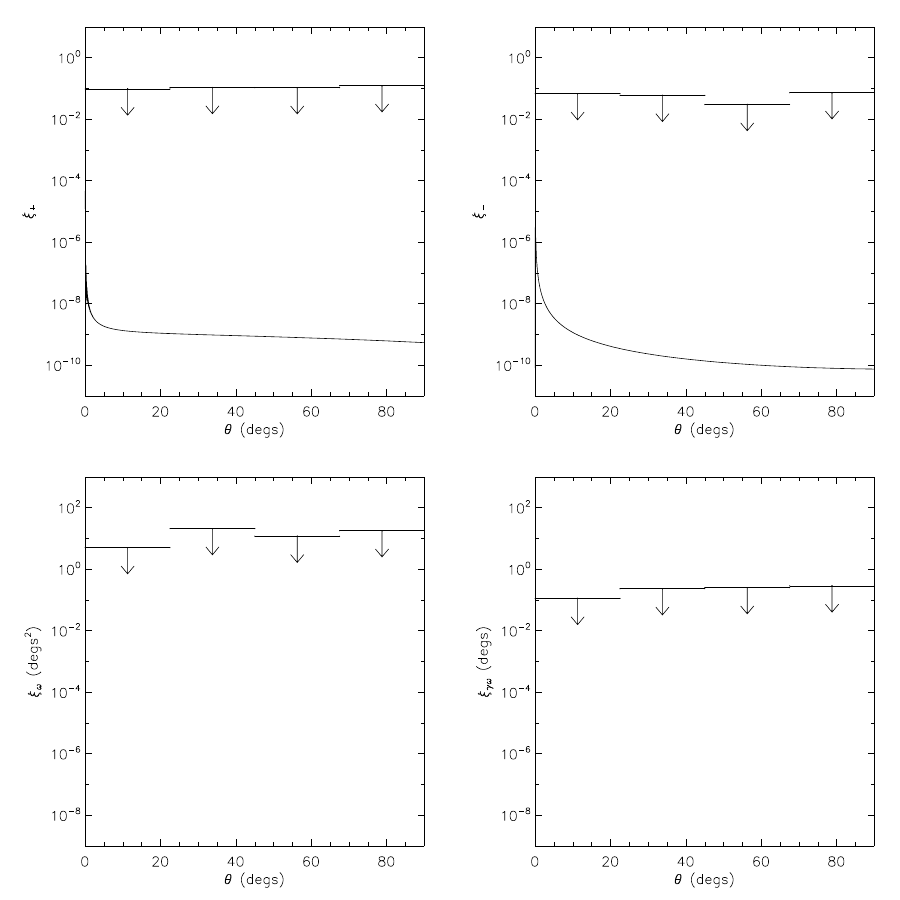}
\caption{Upper (95\% confidence) limits for the shear and rotation 2PCFs calculated using equation (\ref{eq:shear_2pf}). The black curves in the top two panels show the theoretical 2PCFs calculated assuming the Planck best-fit parameters and the spline fit to the source redshift distribution shown in Figure \ref{fig:redshift}.}
\label{fig:correls}
\end{minipage}
\end{figure*}

\begin{figure}
\includegraphics{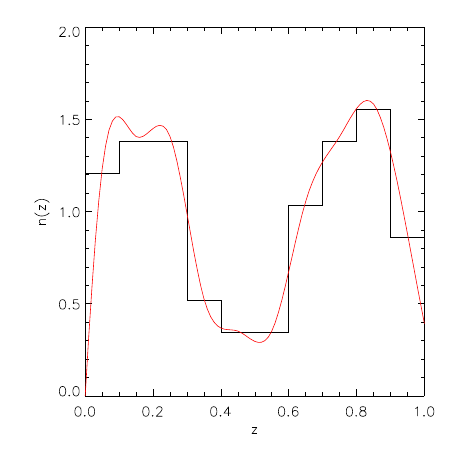}
\caption{The redshift distribution of the full sample of 58 sources. The red curve is a cubic spline fit, which was used to calculate the theoretical shear 2PCFs shown in Figure \ref{fig:correls}.}
\label{fig:redshift}
\end{figure}

The results for the 2PCFs are shown in Figure \ref{fig:correls}. The black curves in the top four panels are theoretical predictions of the shear 2PCFs calculated with {\tt CosmoSIS}\footnote{\href{https://bitbucket.org/joezuntz/cosmosis/wiki/Home}{https://bitbucket.org/joezuntz/cosmosis/wiki/Home}} using the small angle approximation, and hence the theoretical spectra are only accurate to $\sim$$10^{\circ}$ but serve to give an indication of the constraining power of our method for the small sample of sources used. We assume a $\Lambda$CDM cosmology with the Planck best-fit parameters $\Omega_{\mathrm{m}}=0.3089$, $h=0.6774$, $\Omega_{\mathrm{b}}=0.0486$, $n_{\mathrm{s}}=0.9667$, $A_{\mathrm{s}}=2.441\times10^{-9}$, and $\sigma_8=0.8159$ \citep{ade15}. For the redshift distribution, we used a cubic spline fit to the distribution of redshifts for the full sample of 58 sources, shown in Figure \ref{fig:redshift}.

From Figure \ref{fig:correls}, we see that the 95\% confidence upper limits on the shear correlation functions are approximately seven orders of magnitude larger than the predicted correlation functions for this small sample of sources. Given that we have only used 30 sources for this analysis, this level of constraining power comes as no surprise, and for future surveys where a much larger number of sources will be available, such as those performed with the SKA, we expect that it should be possible to recover a clear detection of the signal; this is discussed in more detail in the following section. One exciting result that we see in Figure \ref{fig:correls} is that the average rotation signal has been constrained to $\sim$$1^{\circ}$, which is competetive with current constraints using CMB observations (see e.g. \cite{kaufman16}). Using future radio surveys, we expect that contraints on the rotation signal may be increased by at least an order of magnitude. 

\section{Future observations}
\label{sec:future}
In the previous section, we placed constraints on the shear and rotation 2PCFs using only 30 resolved radio sources. Future high resolution radio surveys will provide a much higher number of well resolved sources, and in this section we investigate the number of sources that would be required to make a clear detection of shear power. To achieve this, we define a figure of merit (FOM) based on the total signal-to-noise of the power spectrum and assume a toy model based on a SKA\footnote{\url{https://www.skatelescope.org}}-like survey.
   
For SKA2, the total sky coverage is expected to be $\sim$30,000$\,\mathrm{deg}^2$, giving a fractional sky coverage of $f_{\mathrm{sky}}\sim$0.7. For this fractional sky coverage, it should be possible to probe the shear power spectrum down to multipoles of $l_{\mathrm{min}}=2$. Assuming that $l_{\mathrm{max}}=2N_{\mathrm{side}}$ (using {\tt HEALPix}\footnote{\href{http://healpix.sourceforge.net/index.php}{http://healpix.sourceforge.net/index.php}} terminology) and that we have one source in each pixel, we have
\begin{equation}\label{eq:lmax_nsource1}
\frac{N_{\mathrm{source}}}{N_{\mathrm{pix}}}=\frac{N_{\mathrm{source}}}{12N^2_{\mathrm{side}}f_{\mathrm{sky}}}=\frac{N_{\mathrm{source}}}{3l_{\mathrm{max}}^2f_{\mathrm{sky}}}=1,
\end{equation}
so that
\begin{equation}\label{eq:lmax_nsource}
l_{\mathrm{max}}=\sqrt{\frac{N_{\mathrm{source}}}{3f_{\mathrm{sky}}}}.
\end{equation}

The noise power spectrum is given by
\begin{equation}\label{eq:n_l}
N_l=\frac{4\pi\sigma^2_{\mathrm{rms}}}{N_{\mathrm{source}}/f_{\mathrm{sky}}},
\end{equation}
where $\sigma_{\mathrm{rms}}$ is the r.m.s. error of the shear estimates for each pixel. As future surveys should provide a high number of high resolution radio sources, for the following discussion, we set $\sigma_{\mathrm{rms}}=8.77\times10^{-2}$, which is the minimum error for the shear estimates given in table \ref{table:derived}. The errors on the estimates of the $C_l$s are then
\begin{equation}\label{eq:error_cl}
\sigma^2_{\hat{C}_l}=\frac{2}{\left(2l+1\right)f_{\mathrm{sky}}}\left(C_l+N_l\right)^2.
\end{equation}

In order to place constraints on the number of galaxies required to make a detection of the shear power spectrum, we bin all of the estimated $C_l$s into one bin. Assuming that the errors on the mean $C_l$ in the bin are Gaussian distributed, we can then estimate how many sources are required to make a detection of the total shear power at a given confidence level. To do this, we define a FOM for the $C_l$ estimates as
\begin{align}\label{eq:FOM}
\mathrm{FOM}=\frac{\frac{1}{N}\sum_{l=l_{\mathrm{min}}}^{l_{\mathrm{max}}}C_l}{\sigma_{<\hat{C}_l>}},
\end{align}
where $\sigma_{<\hat{C}_l>}$ is the error on the mean estimated $C_l$ in the single bin and $l_{\mathrm{max}}$ is a function of the number of sources (equation (\ref{eq:lmax_nsource})). Equation (\ref{eq:FOM}) can be written in terms of $\sigma^2_{\hat{C}_l}$ as
\begin{align}
\mathrm{FOM}=\frac{\sum_{l=l_{\mathrm{min}}}^{l_{\mathrm{max}}}C_l}{\sqrt{\sum_{l=l_{\mathrm{min}}}^{l_{\mathrm{max}}}\sigma^2_{\hat{C}_l}}}.
\end{align}

\begin{table}
\centering
\begin{tabular}{|c|c|c|c|}
\hline
FOM & $N_{\mathrm{source}}$ & $l_{\max}$ & $\sigma_{\left<\omega\right>}$ ($\times10^{-3}\,\mathrm{degs}$) \\ [0.1ex]
\hline
1 & $8.61\times10^4$ & 198 & 7.16 \\ [0.1ex]
3 & $2.49\times10^5$ & 337 & 4.21 \\ [0.1ex]
5 & $4.22\times10^5$ & 439 & 3.23 \\ [0.1ex]
10 & $9.25\times10^5$ & 651 & 2.18 \\ [0.1ex]
\hline
\end{tabular}
\caption{The required $N_{\mathrm{source}}$ to achieve a detection of shear power for a SKA-like survey with FOM = 1, 3, 5, and 10. In the last column, we also present the expected error on a global rotation, $\left<\omega\right>$, for this number of sources.}
\label{table:FOM}
\end{table}

We calculated the theoretical shear power spectrum for a hypothetical SKA2-like radio survey using {\tt CosmoSIS}. We assumed a $\Lambda$CDM model with the Planck best-fit cosmological parameters discussed in the previous section. We also assumed the redshift distribution fitted to the Combined EIS-NVSS Survey Of Radio Sources (CENSORS) \citep{best03, brookes08, caballero13}, which is given as
\begin{equation}
 \frac{\mathrm{d}n}{\mathrm{d}z}=A\left(\frac{z}{z_0}\right)^{\alpha}e^{-\alpha z/z_0},
\end{equation}
where $z_0$ is the mode of the distribution, $\alpha$ is a shape parameter, and $A$ is a normalization constant. The best-fit parameters to the CENSORS data are $z_0=0.53^{+0.11}_{-0.13}$ and $\alpha=0.81^{+0.34}_{-0.32}$. We then calculated the FOM, given in equation (\ref{eq:FOM}), using equation (\ref{eq:error_cl}) with $f_{\mathrm{sky}}=0.7$ and $\sigma_{\mathrm{rms}}=8.77\times10^{-2}$, as discussed above.

\begin{figure}
\includegraphics{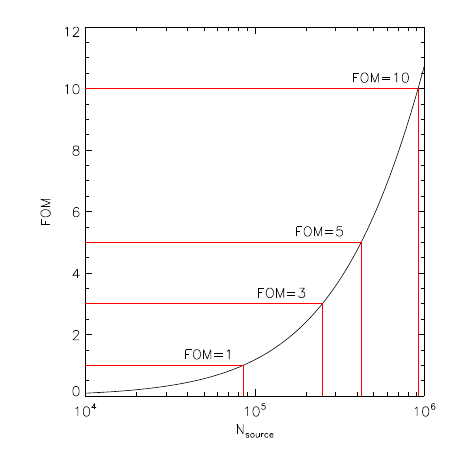}
\caption{The FOM for an SKA2-like survey as a function of $N_{\mathrm{source}}$. The numerical results are given in table \ref{table:FOM}.}
\label{fig:n_ska}
\end{figure}

The results of this test are shown in Figure (\ref{fig:n_ska}). We have indicated the positions where FOM = 1, 3, 5, and 10, corresponding to a $1\sigma$, $3\sigma$, $5\sigma$, and $10\sigma$ detection of shear power respectively. The numerical values are given in table \ref{table:FOM}. We find that for $N_{\mathrm{source}}$$\sim$$10^{5}$ the shear power is approximately equal to the uncertainty, which is expected to be dominated by errors on the shear estimates. A $5\sigma$ detection of shear power is expected for $N_{\mathrm{source}}$$\sim$$5\times10^{5}$. We also see that the maximum multipole which may be probed is $l_{\mathrm{max}}$$\sim$$200$ for $N_{\mathrm{source}}$$\sim$$10^{5}$ and increases to $l_{\mathrm{max}}$$\sim$$700$ for $N_{\mathrm{source}}$$\sim$$10^{6}$. We expect that these numbers of well resolved radio sources should be achievable with future surveys.

In the last column of table \ref{table:FOM}, we also show the predicted uncertainty on a global rotation estimate for these source numbers. As with the shear, we have assumed an average error on the rotation estimates equal to the minimum error in table \ref{table:derived}. We see that with these numbers of sources, the errors on $\left<\omega\right>$ are $\sim$$10^{-3}$ degrees. Hence, our estimator has the  potential to provide powerful constraints on a global rotation signal.

\section{Conclusions}
\label{sec:conclude}
{We find that the} polarization position angles in resolved radio sources are correlated with the position angles of the gradient of the total intensity field. We have shown that this correlation is broken in the presence of a lensing signal and/or an overall rotation of the plane of polarization, and we have formulated an estimator that exploits this effect to provide an estimate of the shear and rotation signals.

The estimator has been successfully demonstrated on simulations, where we have shown that for very high signal-to-noise sources and assuming zero astrophysical scatter between the polarization and gradient orientations, the bias in the estimator is negligible. In the presence of a more realistic signal-to-noise level and astrophysical scatter between the orientations, we find that there is a bias in the estimates. We have developed a method which uses the total intensity and polarization maps for a given source to produce calibration simulations in order to mitigate this effect.

Using data observed with the VLA and MERLIN array, we have measured the shear and rotation signals in the directions of 30 sources and used these estimates to place constraints on the correlation functions. We find that we can constrain the rotation signal to $<$$1^{\circ}$ with this number of sources. This level of constraint is already comparable with constraints from CMB obsevations. Future radio surveys, such as those conducted with the SKA, are expected to deliver a much larger number of higher signal-to-noise sources which should provide a detection of the shear signal, enabling the method to be used to constrain cosmology. We will also be provided with much tighter constraints on the rotation signal, and hence on cosmic birefringence.

The estimate of a cosmological rotation signal was $-2.03^{\circ}\pm0.75^{\circ}$, where an emphasis should placed on the promising size of the error and not upon the magnitude of the estimated signal {since we do not include a correction for Faraday rotation.} In future surveys, where a much greater level of constraint on the rotation signal is expected, the effects of Faraday rotation will become more problematic, and  a map of the rotation measure across a source will be required. However, assuming that the effects of Faraday rotation, and other systematics, can be successfully removed from estimates of the rotation, the level of constraining power on the rotation signal for our approach should be orders of magnitude greater than those currently available.

A clear detection of the shear signal should be possible for future high-precision surveys, which will yield a large number of well resolved sources. One advantage of our approach is that the shear estimates require no assumptions to be made about the intrinsic source shapes. We are therefore insensitive to intrinsic alignment contamination. Also, the high resolution and depth of future surveys should provide many sources with a large number of effective pixels above the signal-to-noise threshold. This could potentially reduce the errors on the shear estimates to well below the level of irreducible shape noise inherent in traditional approaches to weak lensing.

\section*{Acknowledgments}
We thank Paddy Leahy and Martin Hardcastle for providing the data used in this paper. We also thank Ian Browne and Paddy Leahy for useful comments. LW and MLB are grateful to the ERC for support through the award of an
ERC Starting Independent Researcher Grant (EC FP7 grant number 280127). MLB also thanks the STFC for the award of Advanced and
Halliday fellowships (grant number ST/I005129/1). 

\bibliographystyle{mn2e}
\bibliography{ms}

\label{lastpage}

\end{document}